\documentclass[aps,prl,twocolumn,superscriptaddress,nobalancelastpage,longbibliography]{revtex4-2}
\usepackage{amsmath}
\usepackage{amsfonts}
\usepackage{amssymb}
\usepackage{graphicx}
\usepackage{xcolor}
\usepackage[colorlinks=True,citecolor=blue,linkcolor=blue,urlcolor=blue]{hyperref}
\usepackage{hypcap}
\usepackage{braket}
\usepackage{bm}
\usepackage{bbm}
\usepackage[export]{adjustbox}
\usepackage{verbatim}

\definecolor{myBlue}{RGB}{50,119,220}
\definecolor{myOrange}{RGB}{255,127,14}
\definecolor{myGreen}{RGB}{44,160,44}
\definecolor{myRed}{RGB}{214,39,40}
\definecolor{myPurple}{RGB}{148,103,189}

\definecolor{otherBlue}{cmyk}{1, 0.53, 0, 0.1}
\definecolor{otherOrange}{cmyk}{0, 0.72, 1.0, 0.06}

\definecolor{anotherBlue1}{rgb}{0.1271049596, 0.4401845444, 0.7074971164}
\definecolor{anotherBlue2}{rgb}{0.0313725490, 0.2897347174, 0.570319108}
\definecolor{anotherBlue3}{rgb}{0.2909803921, 0.5945098039, 0.7890196078}
\definecolor{anotherBlue4}{rgb}{0.4980392156, 0.7254133025, 0.8561322568}
\definecolor{anotherGrey1}{rgb}{0.1790080738, 0.1790080738, 0.1790080738}
\definecolor{anotherGrey2}{rgb}{0.0091041906, 0.0091041906, 0.0091041906}
\definecolor{anotherGrey3}{rgb}{0.3713033448, 0.3713033448, 0.3713033448}
\definecolor{anotherGrey4}{rgb}{0.5344098423, 0.5344098423, 0.5344098423}
\definecolor{anotherOrange1}{rgb}{0.8462745098, 0.28069204152, 0.004106113033}

\makeatletter
\def\p@figure{\color{blue}}
\def\p@equation{\color{blue}}

\makeatother

\begin{document}

\title{Magnetic effects of non-magnetic impurities in gapped short-range resonating valence bond spin liquids}

\author{Md Zahid Ansari}
\affiliation{Department of Theoretical Physics, Tata Institiute of Fundamental Research, Mumbai 400 005, India} 
\author{Kedar Damle}
\affiliation{Department of Theoretical Physics, Tata Institiute of Fundamental Research, Mumbai 400 005, India} 
\begin{abstract}
We study the effect of a small density $n_v$ of quenched non-magnetic impurities, {\em i.e.} vacancy disorder, in  gapped short-range resonating valence bond (RVB) spin liquid states and valence bond solid (VBS) states of quantum magnets.  We argue that a large class of short-range RVB liquids are stable at small $n_v$ on the kagome lattice, while the corresponding states on triangular, square, and honeycomb lattices are unstable at any nonzero $n_v$ due to the presence of emergent vacancy-induced local moments.  
In contrast, VBS states are argued to be generically unstable (independent of lattice geometry) at nonzero $n_v$ due to such a local-moment instability. Our arguments rely in part on an analysis of the statistical mechanics of maximally-packed dimer covers of the diluted lattice, and are fully supported by our computational results on $O(N)$ symmetric designer Hamiltonians.

\end{abstract}

\maketitle

The original proposal of Anderson~\cite{Anderson_Materials_Research_Bulletin_1973, Fazekas_Anderson_The_Philosophical_Magazine_1974} for a quantum spin liquid state of insulating magnets used a nearest neighbor resonating valence bond (RVB) picture of the many-body ground state.  In any term of the corresponding wavefunction, each spin $S=1/2$ makes a singlet valence bond with one of its neighbors, and the full quantum state is then a superposition over all possible ways in which this can be done. In more general gapped short-range RVB states,  the valence bonds can extend to further neighbors, but are limited in range by a characteristic scale $\xi_{\rm RVB}$~\cite{Liang_Doucot_Anderson_PRL_1988}.

Although the ground state for the particular case considered originally, namely that of the spin $S=1/2$ Heisenberg antiferromagnet on the triangular lattice, is now known~\cite{Huse_Elser_PRL_1988,Gelfand_Singh_Huse_PRB_1989,Singh_Huse_PRL_1992,Bernu_Lhuillier_Pierre_PRL_1992,
Chubukov_Sachdev_Senthil_JPhysCondMat_1994, Capriotti_Trumper_Adolfo_Sorella_PRL_1999,White_Chernyshev_PRL_2007} to not be of this RVB liquid type, this proposal sparked many of the developments~\cite{Kivelson_Rokhsar_Sethna_PRB_1987, Rokhsar_Kivelson_PRL_1988,
Moessner_Sondhi_PRL_2001, Kalmeyer_Laughlin_PRL_1987,  Affleck_Marston_PRB_1988,Read_Sachdev_NuPhysB_1989, Read_Sachdev_PRL_1991,Sachdev_PRB_1992,Misguich_Lhuillier_Bernu_Waldtmann_PRB_1999} that contributed to our modern understanding of such spin liquid states and their characterization in terms of topological order~\cite{Wen_PRB_1991,Wen_QFTbook_2007,Wen_RevModPhys_2017,Moessner_Moore_TPMbook_2021,Sachdev_QPMbook_2023}.
This theoretical progress has also led to experimental efforts aimed at identifying candidate materials that realize such RVB spin liquid states~\cite{Norman_RevModPhys_2016,Broholm_Cava_Kivelson_Nocera_Norman_Senthil_Science_2020,
Clark_Abdeldaim_Annual_Review_of_Material_Research_2021}. 

Part of the difficulty in reaching a definite conclusion about spin liquid behavior in any  material is that the simplest phenomenological characterization of a spin liquid is a negative one: A spin liquid displays no magnetic order of any kind, nor do the spins form a definite static pattern of singlet valence bonds that breaks the symmetry of the underlying crystal structure, as is the case in a valence-bond solid (VBS) state. It is thus defined by what is {\em not seen} in the corresponding experiments.  

As a result, the experimental search for spin liquids is challenging even if material imperfections and quenched disorder effects are absent, and their presence only adds to the challenge~\cite{Broholm_Cava_Kivelson_Nocera_Norman_Senthil_Science_2020,
Clark_Abdeldaim_Annual_Review_of_Material_Research_2021,Lee_nature_2007,
Vries_Kamenev_Kockelmann_Benitez_Harrison_PRL_2008,Ellen_Vojta_Doert_PRM_2022,
Paddison_Daum_Dun_Ehlers_Liu_Stone_Zhou_Mourigal_Nature_Physics_2017}.
Short-range RVB spin liquids are expected to be stable to weak bond disorder (exchange disorder) while being destabilized by  strong bond disorder, while VBS states are predicted to be unstable even to weak bond disorder~\cite{Kimchi_Nahum_Senthil_PRX_2018}. Here, we focus on another important source of disorder, namely non-magnetic substitutional impurities~\cite{Lee_nature_2007,
Vries_Kamenev_Kockelmann_Benitez_Harrison_PRL_2008,
Ellen_Vojta_Doert_PRM_2022}, which can be modeled theoretically as missing spins or static vacancies, and can serve as a probe of the underlying many-body state~\cite{Sandvik_Dagotto_Scalapino_PRB_1997,
Sachdev_Buragohain_Vojta_Science_1999,
Gregor_Motrunich_PRB_2009,Wang_Sandvik_PRB_2010,
Ghosh_Changlani_Henley_PRB_2015,
Hoglund_Sandvik_Sachdev_PRL_2007,
Kaul_Melko_Metlitski_Sachdev_PRL_2008,
Banerjee_Damle_Alet_PRB_2010,
Sanyal_Banerjee_Damle_PRB_2011,
Banerjee_Damle_Alet_PRB_2011}.
 We argue that local moments form in the vicinity of each vacancy  in VBS states even when the vacancies are isolated, {\em i.e.} well-separated from each other, while such local moments arise in  gapped short-range RVB spin liquids {\em only} if the maximum matchings (maximally-packed dimer covers) of the diluted lattice have a nonzero number of monomers, which is typically not the case when the vacancies are isolated~\cite{Aldred_Anstee_Locke_Discrete_Mathematics_2007,Anstee_Blackman_Yang_Discrete_Mathematics_2011,
 Tseng_Anstee_2006}.

On the randomly site-diluted kagome lattice, we find at low dilution that the largest connected ``bulk'' component of the lattice hosts at most one such vacancy-induced monomer or local moment; the bulk density of such local moments is thus zero in this regime. 
  In contrast, on randomly site-diluted triangular, square, and honeycomb lattices, certain vacancy clusters that generically occur with nonzero bulk density lead to a bulk density of such monomers (and hence local moments) situated inside well-demarcated ``${\mathcal R}$-type regions'' whose geometry has been studied previously~\cite{Bhola_Biswas_Islam_Damle_PRX_2022, Bhola_Damle_arXivOct2023} in other contexts. 

We argue that such a bulk density of vacancy-induced emergent local moments represents an instability of the system since the real ground state is then determined by the many-body wavefunction of this system of emergent local moments. Thus, we conclude that gapped short-range RVB liquids are stable at small $n_v$ on the kagome lattice, while such states on the triangular, square, and honeycomb  lattices have a local moment instability at small $n_v$. In contrast, VBS states are always unstable at small $n_v$ independent of lattice geometry.

We first explore the contrasting consequences of site dilution on gapped short-range RVB and VBS states by working within the quantum dimer model (QDM) framework of Rokhsar and Kivelson, where each dimer represents a nearest-neighbor singlet~\cite{Rokhsar_Kivelson_PRL_1988, Moessner_Sondhi_PRL_2001}. Although this framework does not make any explicit reference to further neighbor valence bonds, the effects of matrix elements to such states are encoded via additional terms in the quantum dimer model Hamiltonian~\cite{Rokhsar_Kivelson_PRL_1988}. 
\begin{figure}
\begin{tabular}{cc}
\includegraphics[width=0.5\columnwidth]{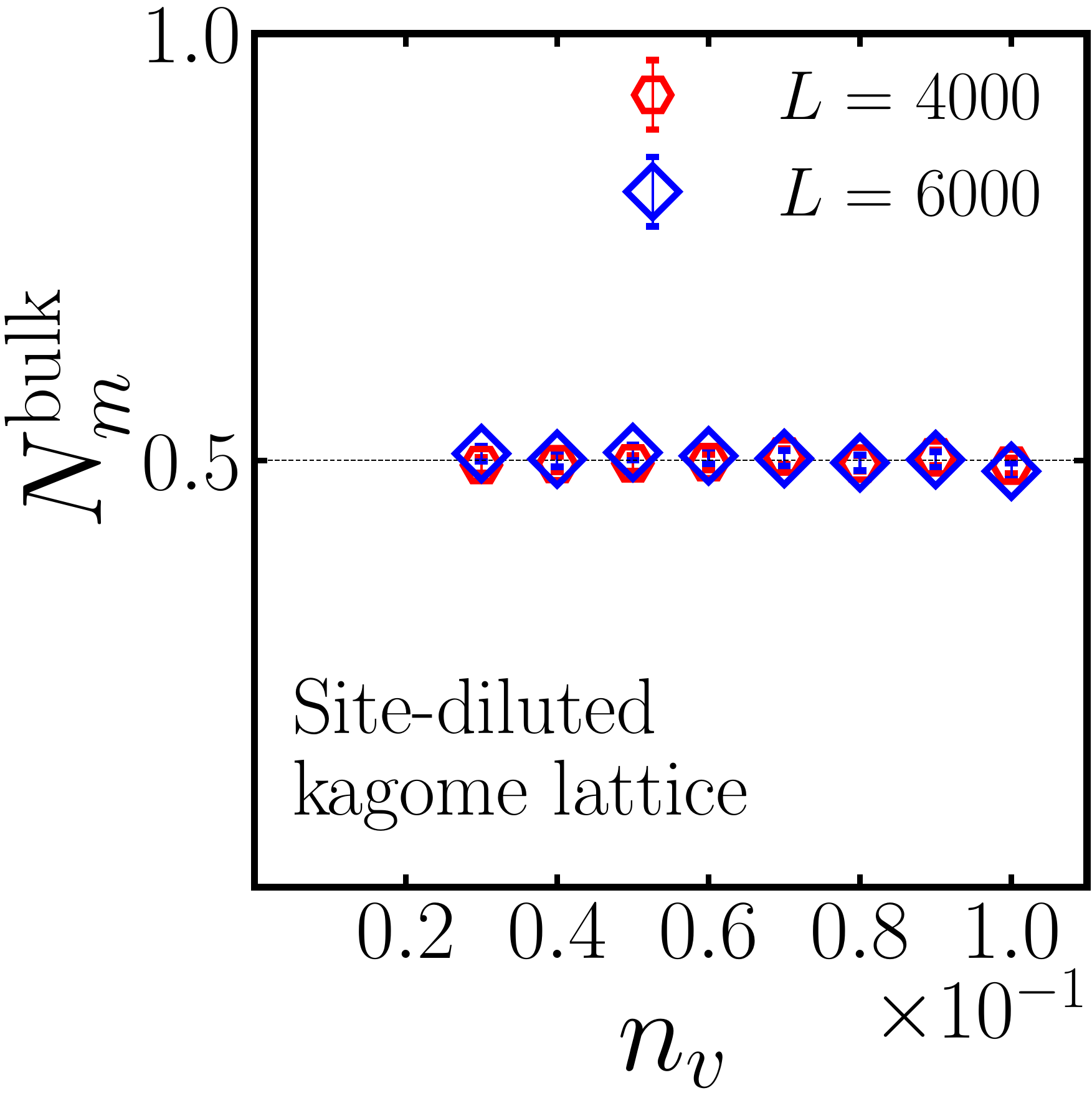} & \includegraphics[width=0.5\columnwidth]{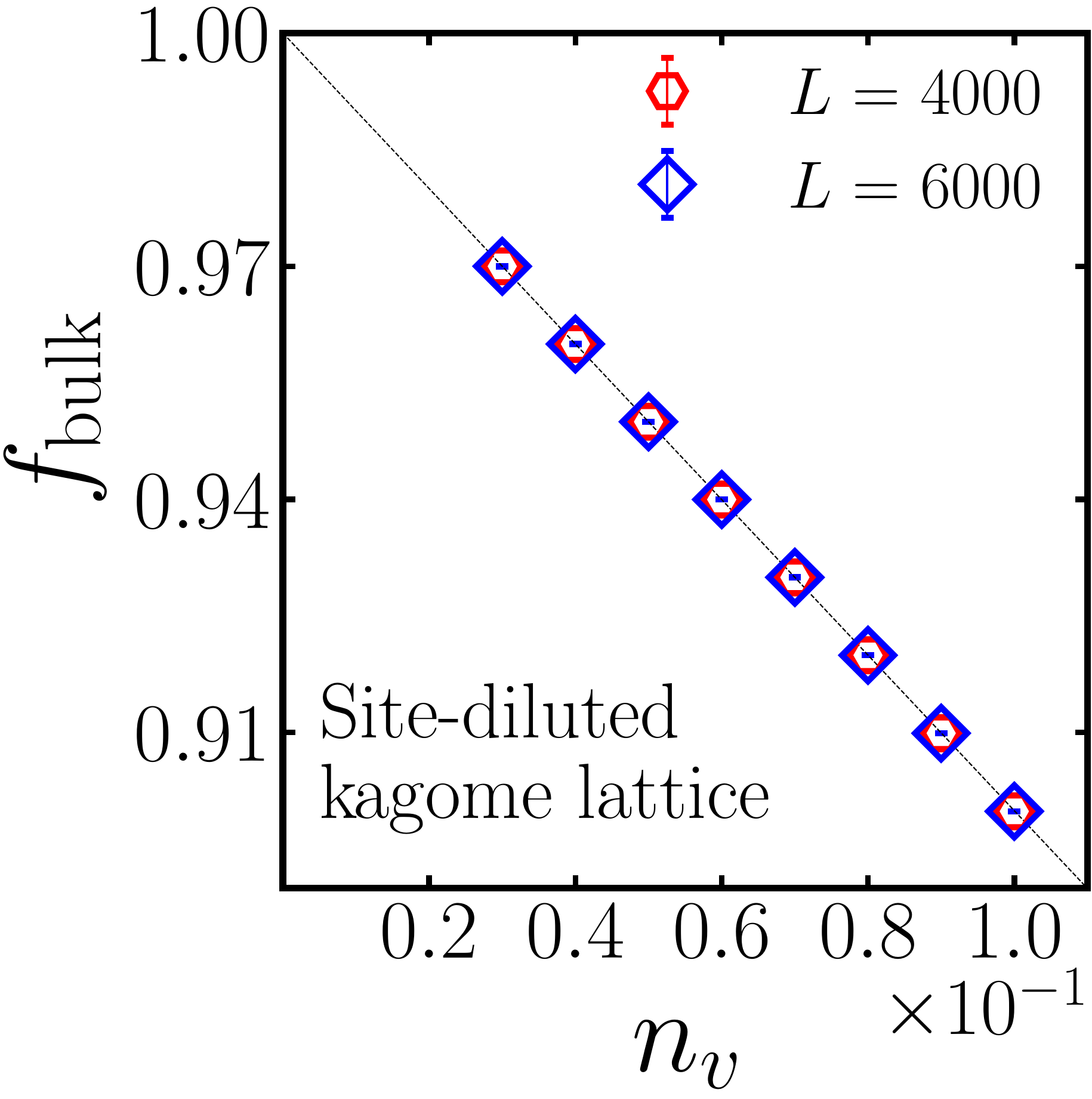} 
\end{tabular}
\caption{ Left panel: Maximum matchings of the largest connected``bulk'' component of a site-diluted kagome lattice have mean (disorder averaged) monomer number $N_m^{\rm bulk} = 0.5$ (within statistical errors), independent of the vacancy density $n_v$. More generally, we have checked that the number of monomers in any connected component is always equal to $1$ ($0$) if the number of sites in it is odd (even). Right panel: At small $n_v$ deep within the geometrically percolated phase of the diluted lattice, this ``bulk'' component  has a very large fraction $f_{\rm bulk}$ of the sites of the diluted lattice.}
\label{fig:kagomeMonomerDensity}
\end{figure}

Within this framework, it is clear that the effect of vacancies depends crucially on whether the diluted lattice has  perfect matchings (perfect dimer covers). If it does, then it typically has exponentially (in the lattice size) many such perfect matchings due to the possibility of local rearrangements of dimers. If it does not have perfect matchings, then a minimum number of sites have to be left unmatched in any {\em  maximum matching} of the lattice, and the unmatched sites host ``monomers'' of the maximally-packed dimer cover. In such cases, there are typically exponentially many such maximum matchings.  Since such an unmatched site is associated with a free spin, these monomers necessarily correspond to emergent vacancy-induced local moments, independent of whether the ground state is an RVB liquid or a VBS state.
\begin{figure}
\begin{tabular}{cc}
 \includegraphics[width=0.5\columnwidth]{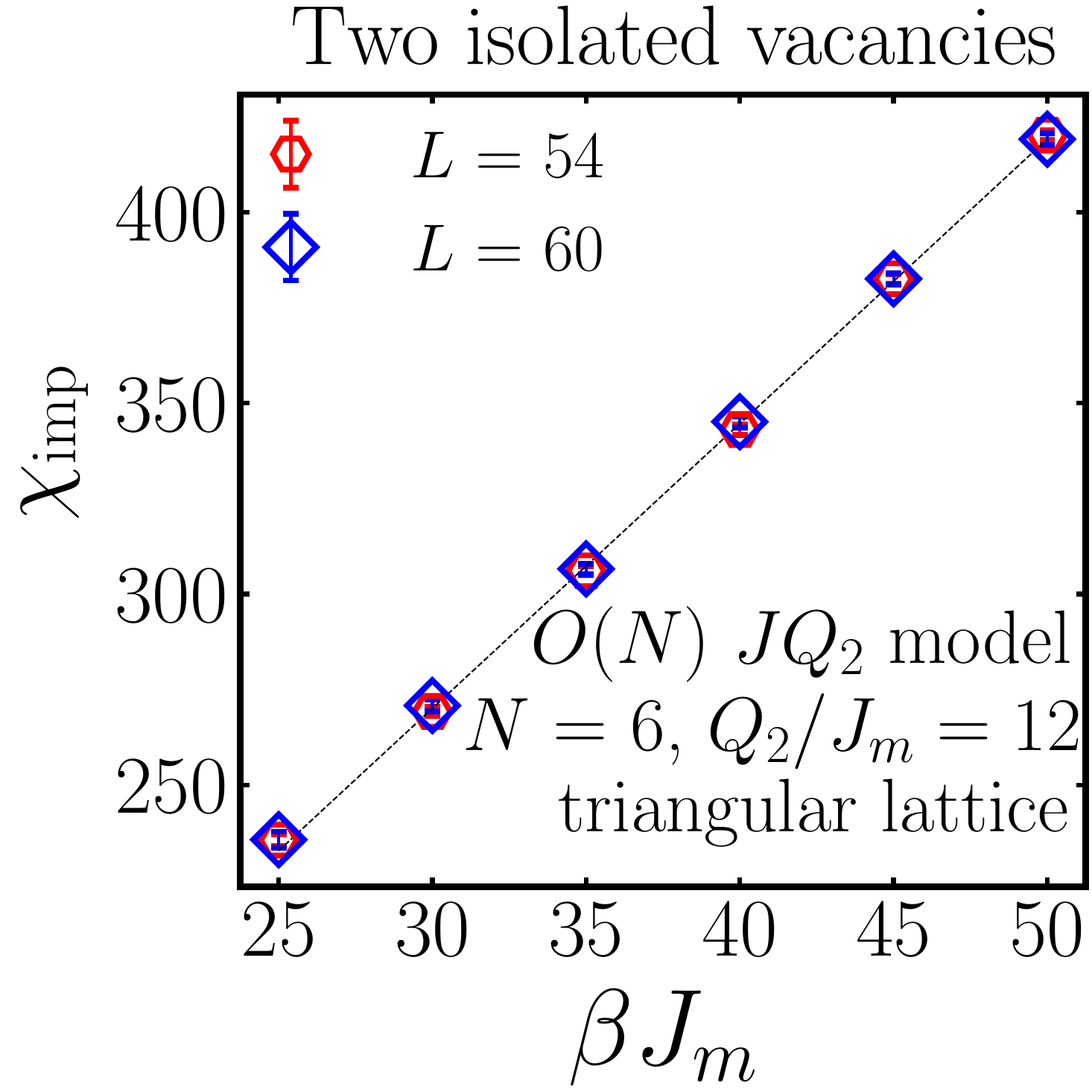} & \includegraphics[width=0.5\columnwidth]{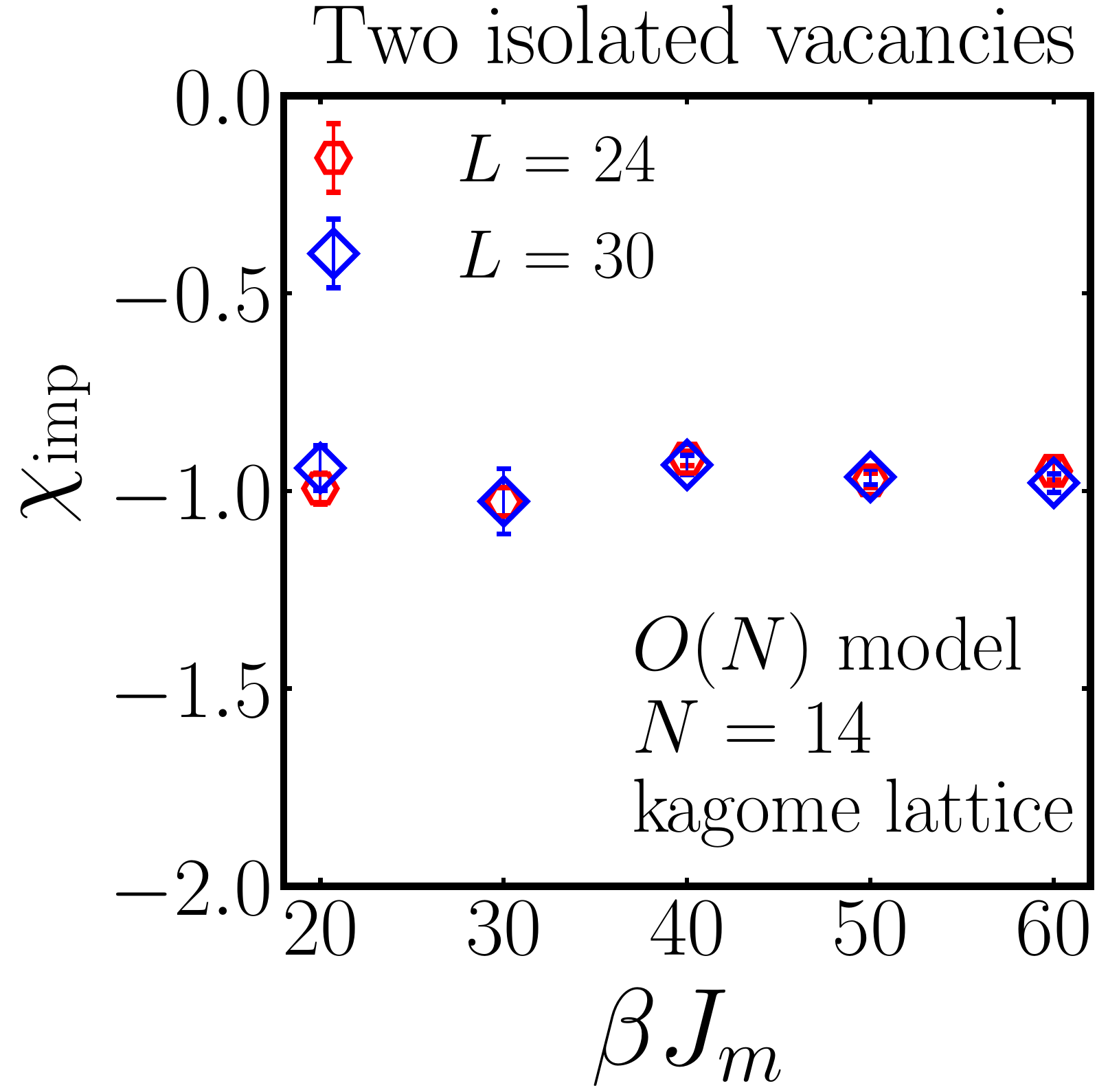}
\end{tabular}
\caption{Left panel: In the triangular lattice $O(N)$ model for values of $N$ in the VBS phase, two isolated vacancies separated by the maximal distance of $L/2$ on an $L\times L$ lattice with $L^2$ unit cells and periodic boundary conditions give rise to a Curie tail, $\chi_{\rm imp} \propto 1/T \equiv \beta$, in the impurity susceptibility $\chi_{\rm imp}$ defined in the main text.  The straight line is a guide to the eye. Right panel: In the kagome case, there is no such Curie tail for values of $N$ in the short-range RVB phase.}
\label{fig:Isolatedvacancies}
\end{figure}

Our key point is this: In a gapped short-range RVB liquid state, there is no energetic preference for any particular arrangement of the singlet valence bonds. Therefore, as long as the diluted lattice has exponentially many perfect dimer covers to form a resonating valence bond state, the system is expected to remain in such an RVB state even at nonzero $n_v$. Thus, vacancies induce the formation of emergent local moments in such RVB states {\em only if} there are monomers in the maximum matchings of the diluted lattice. In contrast, when the state has VBS order, there is a preferred ordered arrangement of valence bonds. This cannot be maintained on a diluted lattice even if it has perfect matchings: Consider for instance a triangular or square lattice with just two isolated vacancies far apart. Although perfect matchings are possible in this case~\cite{Aldred_Anstee_Locke_Discrete_Mathematics_2007,Anstee_Blackman_Yang_Discrete_Mathematics_2011,
 Tseng_Anstee_2006}, any such singlet state corresponding to a fully-packed arrangement of dimers necessarily leads to a domain wall (in the VBS order parameter) connecting the two vacancy locations~\cite{Levin_Senthil_PRB_2004, Senthil_Balents_Sachdev_Vishwanath_Journal_Physical_Society_of_Japan_2005}. The resulting domain wall energy cost (which scales with its length) makes it energetically favourable to eliminate such domain walls by keeping the VBS order intact except in the immediate vicinity of each vacancy. This is achieved by the formation of local moments that are spread out on sites adjacent to each vacancy. Thus, in a VBS ordered system, individual vacancies seed local moments in their vicinity even when they are well-separated from each other and perfect dimer covers exist. Clearly, these conclusions apply equally well to RVB and VBS states of bipartite as well as non-bipartite quantum magnets.

Although this argument seems to rely very crucially on the nearest-neighbor nature of singlet valence bonds in the ground state wavefunction, we nevertheless expect the conclusions to be valid more generally for gapped short-range RVB or VBS states. Indeed, the same conclusions also follow (without relying on any nearest-neighbor singlet wavefunctions) within a $1/N$ expansion approach~\cite{Affleck_PRL_1985,Read_Sachdev_NuPhysB_1989,Kaul_PRL_2015,Block_DEmidio_Kaul_PRB_2020} to $SU(N)$ and $SO(N)$ symmetric generalizations~\cite{Affleck_PRL_1985,Harada_Kawashima_Troyer_PRL_2003,Kawashima_Naoki_Tanabe_PRL_2007,Beach_Alet_Mambrini_Capponi_PRB_2009, Lou_Sandvik_Kawashima_PRB_2009,Kaul_Sandvik_PRL_2012, Block_Melko_Kaul_PRL_2013, Kaul_PRL_2015,Kaul_PRL_2015, Block_DEmidio_Kaul_PRB_2020, Kundu_Desai_Damle_arXivSep2023}
 that have been useful in previous studies of the competition between Neel, VBS and RVB states of quantum magnets~\cite{Harada_Kawashima_Troyer_PRL_2003,Kawashima_Naoki_Tanabe_PRL_2007,Beach_Alet_Mambrini_Capponi_PRB_2009, Lou_Sandvik_Kawashima_PRB_2009,Kaul_Sandvik_PRL_2012,Block_Melko_Kaul_PRL_2013, Kaul_PRL_2015,Kaul_PRL_2015, Block_DEmidio_Kaul_PRB_2020, Kundu_Desai_Damle_arXivSep2023, Sandvik_PRL_2007, Melko_Kaul_PRL_2008, Sandvik_PRL_2010, Sen_Sandvik_PRB_2010, Banerjee_Damle_Paramekanti_PRB_2011, Pujari_Damle_Alet_PRL_2013, Pujari_Alet_Damle_PRB_2015, Iaizzi_Damle_Sandvik_PRB_2017, Iaizzi_Damle_Sandvik_PRB_2018}
. These have a Hamiltonian comprised of nearest-neighbor singlet projectors:
\begin{eqnarray}
\!\!\!\!H  \!\!&=& \!\!- \frac{J_m}{N}\sum_{\langle r_1 r_2\rangle} \sum_{\alpha, \beta} |\alpha\rangle_{r_1} |\alpha\rangle_{r_2} \langle \beta|_{r_1} \langle \beta|_{r_2} + \cdots  \; ,
\label{eq:HSU(N)}
\end{eqnarray}
where $\langle r_1 r_2 \rangle$ denotes nearest-neighbor links connecting adjacent sites, the ``color'' indices $\alpha$ and $\beta$ denote $N$ possible states of the ``spins'', and the ellipses denote possible additional multispin interactions acting on all the spins of a single plaquette or groups of adjacent plaquettes of the lattice. On a nonbipartite lattice, $H$ has global $O(N)$ symmetry, with the colors $\alpha$ and $\beta$ transforming in the fundamental representation of $O(N)$~\cite{Kaul_PRL_2015,Block_DEmidio_Kaul_PRB_2020}. On a bipartite lattice, $r_1$ and $r_2$ always belong to opposite sublattices, and $H$ has enhanced global $SU(N)$ symmetry~\cite{Affleck_PRL_1985,Read_Sachdev_NuPhysB_1989}, with the colors on $A$ ($B$) sublattice sites transforming in the fundamental (complex conjugate of the fundamental) representation of $SU(N)$. [In both cases, the additional terms represented by ellipses respect the corresponding symmetry.] 

Put another way, $H$ always commutes with all the hermitean antisymmetric  generators ${\mathcal A}_{\alpha \beta}^{\rm tot} = \sum_r {\mathcal A}_{\alpha \beta}(r)$ ($\alpha < \beta$) of global $O(N)$ transformations of the colors. Additionally, on bipartite lattices, it also commutes with the symmetric generators ${\mathcal S}_{\alpha \beta}^{\rm tot} = \sum_r (-1)^r {\mathcal S}_{\alpha \beta}(r)$ ($\alpha < \beta$) and the diagonal generators ${\mathcal Q}_{\alpha \alpha}^{\rm tot} = \sum_r (-1)^r {\mathcal Q}_{\alpha \alpha}(r)$ ($\alpha = 1, 2 \dots N-1$, and no sum over the repeated index implied) that enlarge the global symmetry group to $SU(N)$. Here, $(-1)^r = +1$ ($(-1)^r = -1$) for $r$ belonging to the A (B) sublattice, and
\begin{eqnarray}
{\mathcal A}_{\alpha \beta}(r) &=& -i(|\alpha\rangle_r \langle \beta|_r - |\beta\rangle_r \langle \alpha|_r) \; \;  \forall \; \; {\rm pairs} \; \;  \alpha < \beta  \nonumber \\
{\mathcal S}_{\alpha \beta}(r) &=& (|\alpha\rangle_r \langle \beta|_r + |\beta\rangle_r \langle \alpha|_r) \; \; \forall \; \; {\rm pairs} \; \; \alpha < \beta \nonumber \\
{\mathcal Q}_{\alpha \alpha}(r) &=& (|\alpha\rangle_r \langle \alpha|_r -1/N)  \; \; \forall \; \; \alpha = 1 \dots N-1 
\end{eqnarray}
\begin{figure}
\begin{tabular}{cc}
		\includegraphics[width=0.5\columnwidth]{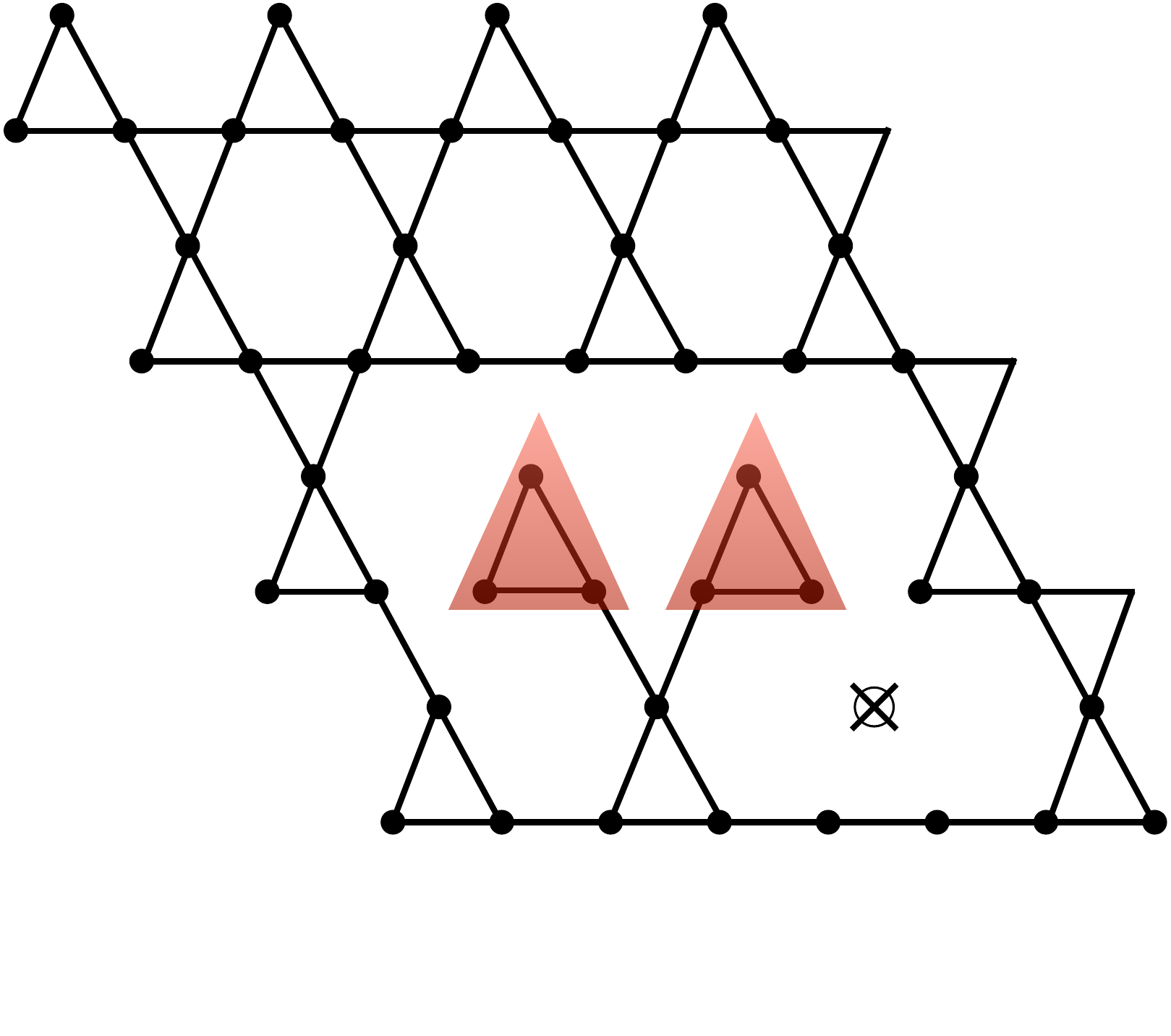} &
		\includegraphics[width= 0.5\columnwidth]{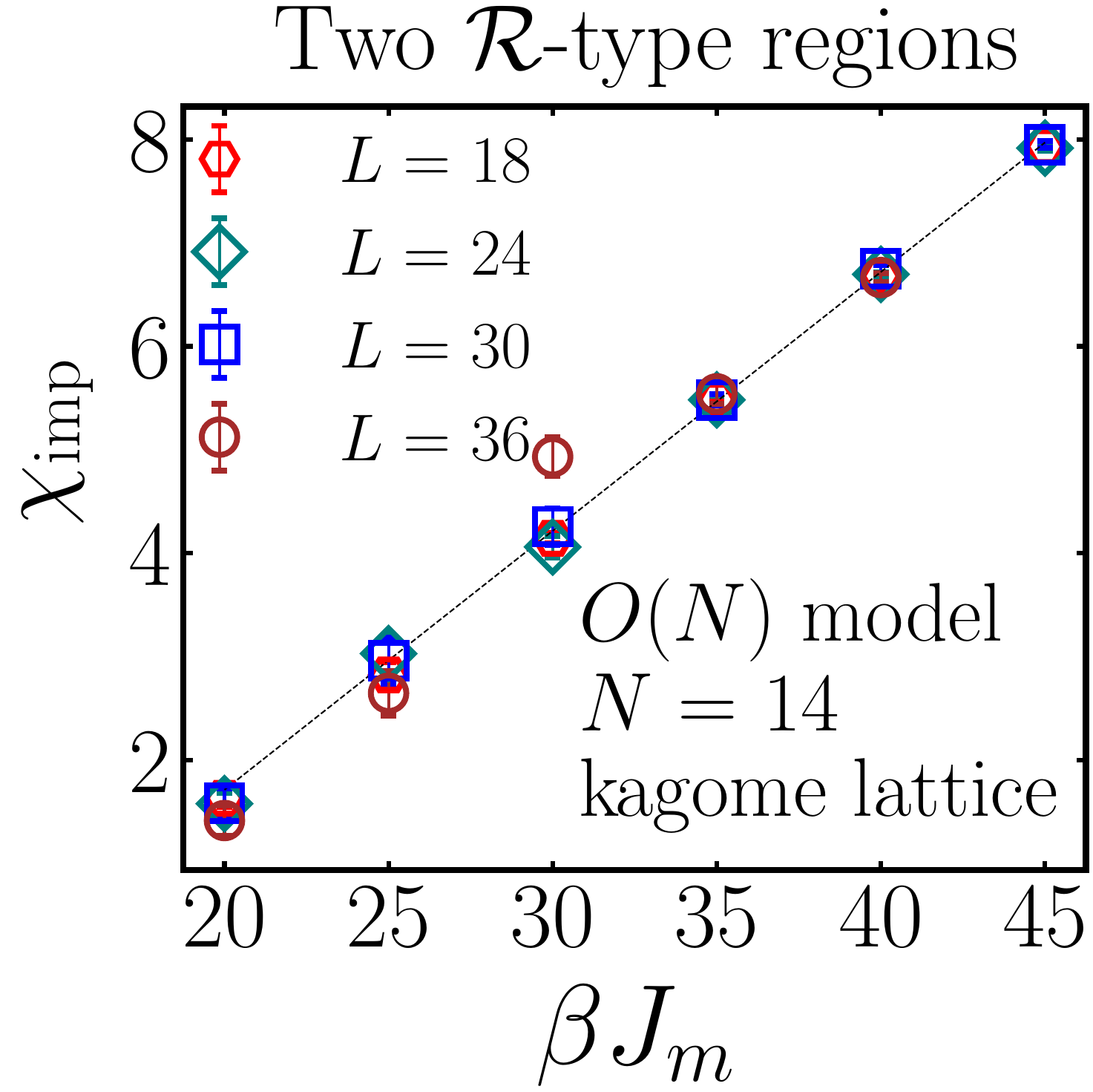}
		\end{tabular}		
	\caption{Left panel:  A small ${\mathcal R}$-type region connected to the bulk of the lattice is  constructed by deleting a set of bonds in the vicinity of a single vacancy. Any maximum matching of the kagome lattice has a single monomer trapped inside it.  Right panel: Impurity susceptibility $\rm \chi_{imp}$ (defined in the  text) of the kagome lattice $O(N)$ model due to two such ${\mathcal R}$-type regions separated by $L/2$ in an otherwise pure $L \times L$ kagome lattice with $L^2$ unit cells and periodic boundary conditions shows clear evidence of a Curie tail $\chi_{\rm imp} \propto 1/T \equiv \beta$ for $N$ in the RVB liquid phase. The straight line is a guide to the eye.
}
	\label{fig:kagomeRtyperegions}
\end{figure}

In the large-$N$ limit without disorder, each perfect dimer cover of the lattice gives a degenerate ground state in both cases, with the dimers representing $SU(N)$ or $O(N)$ singlets. At leading order in $1/N$, the low-energy physics is then controlled by an effective quantum dimer model Hamiltonian~\cite{Affleck_PRL_1985,Read_Sachdev_NuPhysB_1989,Kaul_PRL_2015,Block_DEmidio_Kaul_PRB_2020}.
Generalizing to the diluted case, we see that the ground state degeneracy at $N=\infty$ now corresponds either to fully-packed dimer covers of the lattice, or to the maximally-packed dimer covers if no perfect matchings exist. In the latter case, the monomers of a maximum matching correspond to free $SU(N)$ or $O(N)$ spins. At leading order in $1/N$, one obtains an effective Hamiltonian that acts within this low-energy subspace. This has the form of a quantum monomer-dimer model in the general case, with ring exchange and monomer hopping terms acting within the ensemble of maximum matchings. At higher orders, additional terms are generated, which act on groups of contiguous bonds and plaquettes. From this structure of the low-energy theory, we see that our previous conclusions regarding vacancy-induced local moments in VBS and RVB states apply to these $SU(N)$ and $O(N)$ generalizations at a finite $N$ so long as the corresponding ground state is accessible perturbatively in $1/N$. 

In addition, we see that the perturbatively generated interactions $J_{\rm eff}$ between these emergent local moments only act within each connected component of a diluted lattice.  Due to the gapped nature of the parent state, these effective interaction $J_{\rm eff}$ are also expected to be parametrically smaller  than the microscopic exchange coupling $J_{ m}$ both in the gapped VBS case, and in the gapped short-range RVB case.
Although the form of these interactions is expected to depend sensitively on the microscopic Hamiltonian, the mere fact of their existence implies that a nonzero bulk density of such moments generically signals an instability of the parent state independent of these details~\cite{Kimchi_Nahum_Senthil_PRX_2018}. This is because the long-distance structure of the ground state is now controlled by the many-body wavefunction of this system of emergent local moments.  Clearly, such a bulk density of emergent local moments must also give rise to a vacancy-induced contribution $\chi_{\rm imp} \propto 1/T$ to the thermodynamic susceptibility at low but not too low temperatures in the range $J_{\rm eff} \ll T \ll J_{ m}$. This serves as a diagnostic for the corresponding local moment instability. 

Putting all these arguments together, we are thus led to conclude: short-range VBS states always have a local moment instability at small $n_v$, while short-range RVB states are unstable only if the dilution leads to a bulk density of monomers in the maximum matchings of the diluted lattice.
With this in mind, we turn our attention to the monomers in maximum matchings of randomly site-diluted lattices of interest to us. The results of Ref.~\cite{Bhola_Biswas_Islam_Damle_PRX_2022} and Ref.~\cite{Bhola_Damle_arXivOct2023} make it clear that maximum matchings of site-diluted square, honeycomb, and triangular lattices have a nonzero bulk density of monomers.
For the kagome case, we supplement this with a
computational study using the implementation of Edmonds's maximum matching algorithm given in Ref.~\cite{Kececioglu_Pecqueur_MaxMatch_1998}. We find that the situation is dramatically different: Each connected component (of the diluted kagome lattice) with an odd number of sites hosts exactly one monomer, while connected components with an even number of sites have a perfect dimer cover, {\em i.e.} they host no monomers. 

As a result, as the data in Fig.~\ref{fig:kagomeMonomerDensity} shows, the bulk density of monomers vanishes for small $n_v$ ({\em i.e.} below the geometric percolation threshold of the diluted kagome lattice), since the number of sites in the largest connected component of an $L \times L$  diluted kagome lattice scales as $L^2$ in the thermodynamic limit, while the mean number of monomers in this component is $0.5$. Indeed, we find that essentially all the monomers of any maximum matching of the diluted graph are associated with small disconnected fragments of the diluted lattice at small $n_v$, with the fraction of monomers corresponding to isolated single-site clusters approaching unity as $n_v \to 0$. 

In the kagome spin liquid case, we thus expect vacancy disorder to only lead to at most a single local moment on each individual connected component of the disordered lattice, including on the largest ``bulk" component. Being in different fragments of the lattice, these isolated local moments are not expected to interact with each other via the effective couplings $J_{\rm eff}$. Instead, they represent an essentially decoupled population of free spins that co-exist with the bulk spin liquid state. 

Therefore, we conclude that gapped short-range RVB spin liquid states are stable at small $n_v$ on the kagome lattice, but have a vacancy-induced local moment instability at any nonzero $n_v$ on square, honeycomb and triangular lattices.
For a reliable computational test of our arguments and this conclusion, we need examples of model Hamiltonians that exhibit such ground states while being amenable to large-scale sign-free quantum Monte Carlo (QMC) studies. 
When such model systems can be identified, the most direct and straightforward test applicable to both $SU(N)$ and $O(N)$ generalizations would involve studying the effect of vacancies on the susceptibility $\chi^{{\mathcal A}}$ to a uniform field that couples to ${\mathcal A}_{\alpha \beta}^{\rm tot}$ (for any pair $\alpha < \beta$). This would be the analog of the susceptibility to a uniform external field that couples to $S_y^{\rm tot}$ in $SU(2)$ symmetric bipartite systems. However, this susceptibility is not readily measurable in QMC simulations that work in the color ($S^z$) basis, which are however capable of measuring susceptibilities to fields that couple to diagonal operators in the color basis. For  $SU(N)$ symmetric designer Hamiltonians on bipartite lattices, this is not a constraint, since one can equivalently measure the susceptibility $\chi^{{\mathcal Q}}$ to a field that couples to ${\mathcal Q}_{\alpha \alpha}^{\rm tot}$ (for any $\alpha$), since this is equivalent by $SU(N)$ symmetry to $\chi^{{\mathcal A}}$. 

Clearly, this is not an option for the $O(N)$ generalizations on non-bipartite lattices. However, this difficulty is not a serious obstacle for the following reason: Such emergent local moments are expected to be essentially free in a broad temperature range $J_{\rm eff} \ll T \ll J_m$. In the bipartite $SU(N)$ symmetric case, it is therefore clear that they will lead to a Curie tail not just in $\chi^{{\mathcal Q}}$, but also in the susceptibility $\chi$ to a field  that couples to $n_{\alpha \alpha}^{\rm tot}$ (for any $\alpha$), where $n_{\alpha \alpha}^{\rm tot}$ differs from ${\mathcal Q}_{\rm \alpha \alpha}$ by the absence of the alternating sign $(-1)^r$ in its definition: $n_{\alpha \alpha}^{\rm tot} = \sum_r {\mathcal Q}_{\alpha \alpha}(r)$ [for $N=2$, this is the $\hat{z}$ component of the N\'eel order parameter]. Indeed, the difference between $\chi$ and $\chi^{{\mathcal Q}}$ will become visible only at $T \sim  J_{\rm eff} \ll J_m$. This suggests vacancy-induced local moments will lead to a Curie tail in $\chi$ in the non-bipartite case too, and this can serve as an alternate diagnostic test. [For $N>2$, $n^{\rm tot}_{\alpha \alpha}$ corresponds to the nematic order parameter in such $O(N)$ models].

Examples of designer Hamiltonians with a gapped short-range RVB liquid ground state are in short supply compared to the variety of different models that display VBS ground states~\cite{Harada_Kawashima_Troyer_PRL_2003,Kawashima_Naoki_Tanabe_PRL_2007,Beach_Alet_Mambrini_Capponi_PRB_2009, Lou_Sandvik_Kawashima_PRB_2009,Kaul_Sandvik_PRL_2012,Block_Melko_Kaul_PRL_2013, Kaul_PRL_2015,Kaul_PRL_2015, Block_DEmidio_Kaul_PRB_2020, Kundu_Desai_Damle_arXivSep2023, Sandvik_PRL_2007, Melko_Kaul_PRL_2008, Sandvik_PRL_2010, Sen_Sandvik_PRB_2010, Banerjee_Damle_Paramekanti_PRB_2011, Pujari_Damle_Alet_PRL_2013, Pujari_Alet_Damle_PRB_2015, Iaizzi_Damle_Sandvik_PRB_2017, Iaizzi_Damle_Sandvik_PRB_2018}.
Fortunately for our purposes, recent QMC results have established the presence of a gapped short-range RVB liquid ground state for the nearest-neighbor $O(N)$ projector Hamiltonian $H$ (Eq.~\ref{eq:HSU(N)}) with $N>9$ on the kagome lattice. Therefore, we focus on this $O(N)$ kagome RVB liquid and use large-scale stochastic series expansion (SSE) QMC~\cite{Sandvik_JPhysA_1992, Sandvik_PRB_1999, Syljuasen_Sandvik_PRE_2002, Sandvik_AIP_Conference_2010,Sandvik_Evertz_PRB_2010, Banerjee_Damle_JStatMech_2010,Desai_Pujari_PRB_2021}
to compute  $\chi_{\rm imp}(T) = \chi_{\rm disordered} - \chi_{\rm pure}$, the vacancy-induced change in the static susceptance: 
\begin{equation}
 \chi = \frac{1}{N-1}\sum_{\alpha=1}^{N-1}\sum_{r,r'}\int_{0}^{\beta}\langle \mathcal{Q}_{\alpha \alpha}(r, \tau)\mathcal{Q}_{\alpha \alpha}(r', 0)\rangle d\tau, \label{eq:chist}
\end{equation}
with $\beta =1/T$ being the inverse temperature. 
We contrast it with vacancy effects on the same quantity in the triangular lattice $O(N)$ model with nearest-neighbor couplings $J_m$ (Eq.~\ref{eq:HSU(N)}) and additional four-spin interactions of strength $Q_2$ on four-site plaquettes of the triangular lattice, since this is known to have a VBS ordered ground state for $N>5$ and large enough values of $Q_2/J_m$. On both lattices, we consider two disorder configurations, one consisting of two isolated vacancies at distance $L/2$ in an $L \times L$ sample, and the other consisting of two ${\mathcal R}$-type regions at the same distance, each of which traps one monomer of a maximum matching.

In Fig.~\ref{fig:Isolatedvacancies}, we show our results for $\chi_{\rm imp}$ corresponding to two isolated vacancies separated by $L/2$ in $L \times L$ triangular and kagome lattice $O(N)$ models for values of $N$ in the VBS and short-range RVB phases respectively. These results demonstrate that isolated vacancies give rise to a Curie tail in $\chi_{\rm imp}$ in the VBS case but not in the RVB case, confirming one key part of our argument. In Fig.~\ref{fig:kagomeRtyperegions}, we display a pattern of bond dilution on the kagome lattice that gives rise to an ${\mathcal R}$-type region that traps a single monomer.  In addition, we display our data for $\chi_{\rm imp}$ for an $L \times L$ kagome lattice $O(N)$ model with two such ${\mathcal R}$-type regions separated by $L/2$ for a value of $N$ in the RVB liquid phase. Clearly, such monomer-carrying regions give rise to such a Curie tail even in the RVB case, in contrast to  isolated vacancies, which  do not. This confirms the other key part of our argument. In addition, our results (not shown) confirm that such ${\mathcal R}$-type regions do give rise (exactly as expected) to a Curie tail in $\chi_{\rm imp}$ in the triangular lattice VBS state too. Finally, we note that {\em bond dilution} (modeling missing exchange pathways) is expected to lead to a nonzero monomer density even on the kagome lattice (as is clear from the example in Fig.~\ref{fig:kagomeRtyperegions}). Therefore, we conclude that short-range RVB spin liquid states are generically unstable to bond dilution independent of lattice geometry, although they are expected to be stable~\cite{Kimchi_Nahum_Senthil_PRX_2018} to weak exchange disorder.

{\em Acknowledgements: } 
We thank Leon Balents and Subir Sachdev for a useful discussion, Richard Anstee for an informative correspondence about Refs.~\cite{Aldred_Anstee_Locke_Discrete_Mathematics_2007,Anstee_Blackman_Yang_Discrete_Mathematics_2011,
 Tseng_Anstee_2006} from the mathematical literature, Souvik Kundu and Ritesh Bhola for fruitful collaborations on related work, and departmental system administrators K. Ghadiali and A. Salve for help with cluster related issues. One of us (KD) also gratefully acknowledges the hospitality of IISER Pune and stimulating discussions with Deepak Dhar while finalizing the draft of this manuscript. The work of  MZA was
supported by a graduate fellowship of the DAE, India at the Tata Institute of Fundamental Research (TIFR). KD
was supported at the TIFR by DAE, India, and in part by a J.C. Bose Fellowship (JCB/2020/000047) of SERB,
DST India, and by the Infosys-Chandrasekharan Random Geometry Center (TIFR). All computations were
performed using departmental computational resources of the Department of Theoretical Physics, TIFR and additional resources supported by a J.C. Bose Fellowship grant (JCB/2020/000047).

\bibliography{Reference}

\begin{thebibliography}{76}%
\makeatletter
\providecommand \@ifxundefined [1]{%
 \@ifx{#1\undefined}
}%
\providecommand \@ifnum [1]{%
 \ifnum #1\expandafter \@firstoftwo
 \else \expandafter \@secondoftwo
 \fi
}%
\providecommand \@ifx [1]{%
 \ifx #1\expandafter \@firstoftwo
 \else \expandafter \@secondoftwo
 \fi
}%
\providecommand \natexlab [1]{#1}%
\providecommand \enquote  [1]{``#1''}%
\providecommand \bibnamefont  [1]{#1}%
\providecommand \bibfnamefont [1]{#1}%
\providecommand \citenamefont [1]{#1}%
\providecommand \href@noop [0]{\@secondoftwo}%
\providecommand \href [0]{\begingroup \@sanitize@url \@href}%
\providecommand \@href[1]{\@@startlink{#1}\@@href}%
\providecommand \@@href[1]{\endgroup#1\@@endlink}%
\providecommand \@sanitize@url [0]{\catcode `\\12\catcode `\$12\catcode
  `\&12\catcode `\#12\catcode `\^12\catcode `\_12\catcode `\%12\relax}%
\providecommand \@@startlink[1]{}%
\providecommand \@@endlink[0]{}%
\providecommand \url  [0]{\begingroup\@sanitize@url \@url }%
\providecommand \@url [1]{\endgroup\@href {#1}{\urlprefix }}%
\providecommand \urlprefix  [0]{URL }%
\providecommand \Eprint [0]{\href }%
\providecommand \doibase [0]{https://doi.org/}%
\providecommand \selectlanguage [0]{\@gobble}%
\providecommand \bibinfo  [0]{\@secondoftwo}%
\providecommand \bibfield  [0]{\@secondoftwo}%
\providecommand \translation [1]{[#1]}%
\providecommand \BibitemOpen [0]{}%
\providecommand \bibitemStop [0]{}%
\providecommand \bibitemNoStop [0]{.\EOS\space}%
\providecommand \EOS [0]{\spacefactor3000\relax}%
\providecommand \BibitemShut  [1]{\csname bibitem#1\endcsname}%
\let\auto@bib@innerbib\@empty
\bibitem [{\citenamefont
  {Anderson}(1973)}]{Anderson_Materials_Research_Bulletin_1973}%
  \BibitemOpen
  \bibfield  {author} {\bibinfo {author} {\bibfnamefont {P.}~\bibnamefont
  {Anderson}},\ }\bibfield  {title} {\bibinfo {title} {Resonating valence
  bonds: A new kind of insulator?},\ }\href
  {https://doi.org/https://doi.org/10.1016/0025-5408(73)90167-0} {\bibfield
  {journal} {\bibinfo  {journal} {Materials Research Bulletin}\ }\textbf
  {\bibinfo {volume} {8}},\ \bibinfo {pages} {153} (\bibinfo {year}
  {1973})}\BibitemShut {NoStop}%
\bibitem [{\citenamefont {Fazekas}\ and\ \citenamefont
  {Anderson}(1974)}]{Fazekas_Anderson_The_Philosophical_Magazine_1974}%
  \BibitemOpen
  \bibfield  {author} {\bibinfo {author} {\bibfnamefont {P.}~\bibnamefont
  {Fazekas}}\ and\ \bibinfo {author} {\bibfnamefont {P.~W.}\ \bibnamefont
  {Anderson}},\ }\bibfield  {title} {\bibinfo {title} {On the ground state
  properties of the anisotropic triangular antiferromagnet},\ }\href
  {https://doi.org/10.1080/14786439808206568} {\bibfield  {journal} {\bibinfo
  {journal} {The Philosophical Magazine: A Journal of Theoretical Experimental
  and Applied Physics}\ }\textbf {\bibinfo {volume} {30}},\ \bibinfo {pages}
  {423} (\bibinfo {year} {1974})}\BibitemShut {NoStop}%
\bibitem [{\citenamefont {Liang}\ \emph {et~al.}(1988)\citenamefont {Liang},
  \citenamefont {Doucot},\ and\ \citenamefont
  {Anderson}}]{Liang_Doucot_Anderson_PRL_1988}%
  \BibitemOpen
  \bibfield  {author} {\bibinfo {author} {\bibfnamefont {S.}~\bibnamefont
  {Liang}}, \bibinfo {author} {\bibfnamefont {B.}~\bibnamefont {Doucot}},\ and\
  \bibinfo {author} {\bibfnamefont {P.~W.}\ \bibnamefont {Anderson}},\
  }\bibfield  {title} {\bibinfo {title} {Some new variational
  resonating-valence-bond-type wave functions for the spin-\textonehalf{}
  antiferromagnetic {H}eisenberg model on a square lattice},\ }\href
  {https://doi.org/10.1103/PhysRevLett.61.365} {\bibfield  {journal} {\bibinfo
  {journal} {Phys. Rev. Lett.}\ }\textbf {\bibinfo {volume} {61}},\ \bibinfo
  {pages} {365} (\bibinfo {year} {1988})}\BibitemShut {NoStop}%
\bibitem [{\citenamefont {Huse}\ and\ \citenamefont
  {Elser}(1988)}]{Huse_Elser_PRL_1988}%
  \BibitemOpen
  \bibfield  {author} {\bibinfo {author} {\bibfnamefont {D.~A.}\ \bibnamefont
  {Huse}}\ and\ \bibinfo {author} {\bibfnamefont {V.}~\bibnamefont {Elser}},\
  }\bibfield  {title} {\bibinfo {title} {Simple variational wave functions for
  two-dimensional {H}eisenberg spin-\textonehalf{} antiferromagnets},\ }\href
  {https://doi.org/10.1103/PhysRevLett.60.2531} {\bibfield  {journal} {\bibinfo
   {journal} {Phys. Rev. Lett.}\ }\textbf {\bibinfo {volume} {60}},\ \bibinfo
  {pages} {2531} (\bibinfo {year} {1988})}\BibitemShut {NoStop}%
\bibitem [{\citenamefont {Gelfand}\ \emph {et~al.}(1989)\citenamefont
  {Gelfand}, \citenamefont {Singh},\ and\ \citenamefont
  {Huse}}]{Gelfand_Singh_Huse_PRB_1989}%
  \BibitemOpen
  \bibfield  {author} {\bibinfo {author} {\bibfnamefont {M.~P.}\ \bibnamefont
  {Gelfand}}, \bibinfo {author} {\bibfnamefont {R.~R.~P.}\ \bibnamefont
  {Singh}},\ and\ \bibinfo {author} {\bibfnamefont {D.~A.}\ \bibnamefont
  {Huse}},\ }\bibfield  {title} {\bibinfo {title} {Zero-temperature ordering in
  two-dimensional frustrated quantum {Heisenberg} antiferromagnets},\ }\href
  {https://doi.org/10.1103/PhysRevB.40.10801} {\bibfield  {journal} {\bibinfo
  {journal} {Phys. Rev. B}\ }\textbf {\bibinfo {volume} {40}},\ \bibinfo
  {pages} {10801} (\bibinfo {year} {1989})}\BibitemShut {NoStop}%
\bibitem [{\citenamefont {Singh}\ and\ \citenamefont
  {Huse}(1992)}]{Singh_Huse_PRL_1992}%
  \BibitemOpen
  \bibfield  {author} {\bibinfo {author} {\bibfnamefont {R.~R.~P.}\
  \bibnamefont {Singh}}\ and\ \bibinfo {author} {\bibfnamefont {D.~A.}\
  \bibnamefont {Huse}},\ }\bibfield  {title} {\bibinfo {title}
  {Three-sublattice order in triangular- and kagom\'e-lattice spin-half
  antiferromagnets},\ }\href {https://doi.org/10.1103/PhysRevLett.68.1766}
  {\bibfield  {journal} {\bibinfo  {journal} {Phys. Rev. Lett.}\ }\textbf
  {\bibinfo {volume} {68}},\ \bibinfo {pages} {1766} (\bibinfo {year}
  {1992})}\BibitemShut {NoStop}%
\bibitem [{\citenamefont {Bernu}\ \emph {et~al.}(1992)\citenamefont {Bernu},
  \citenamefont {Lhuillier},\ and\ \citenamefont
  {Pierre}}]{Bernu_Lhuillier_Pierre_PRL_1992}%
  \BibitemOpen
  \bibfield  {author} {\bibinfo {author} {\bibfnamefont {B.}~\bibnamefont
  {Bernu}}, \bibinfo {author} {\bibfnamefont {C.}~\bibnamefont {Lhuillier}},\
  and\ \bibinfo {author} {\bibfnamefont {L.}~\bibnamefont {Pierre}},\
  }\bibfield  {title} {\bibinfo {title} {Signature of {N\'eel} order in exact
  spectra of quantum antiferromagnets on finite lattices},\ }\href
  {https://doi.org/10.1103/PhysRevLett.69.2590} {\bibfield  {journal} {\bibinfo
   {journal} {Phys. Rev. Lett.}\ }\textbf {\bibinfo {volume} {69}},\ \bibinfo
  {pages} {2590} (\bibinfo {year} {1992})}\BibitemShut {NoStop}%
\bibitem [{\citenamefont {Chubukov}\ \emph {et~al.}(1994)\citenamefont
  {Chubukov}, \citenamefont {Sachdev},\ and\ \citenamefont
  {Senthil}}]{Chubukov_Sachdev_Senthil_JPhysCondMat_1994}%
  \BibitemOpen
  \bibfield  {author} {\bibinfo {author} {\bibfnamefont {A.~V.}\ \bibnamefont
  {Chubukov}}, \bibinfo {author} {\bibfnamefont {S.}~\bibnamefont {Sachdev}},\
  and\ \bibinfo {author} {\bibfnamefont {T.}~\bibnamefont {Senthil}},\
  }\bibfield  {title} {\bibinfo {title} {Large-{S} expansion for quantum
  antiferromagnets on a triangular lattice},\ }\href
  {https://doi.org/10.1088/0953-8984/6/42/019} {\bibfield  {journal} {\bibinfo
  {journal} {Journal of Physics: Condensed Matter}\ }\textbf {\bibinfo {volume}
  {6}},\ \bibinfo {pages} {8891} (\bibinfo {year} {1994})}\BibitemShut
  {NoStop}%
\bibitem [{\citenamefont {Capriotti}\ \emph {et~al.}(1999)\citenamefont
  {Capriotti}, \citenamefont {Trumper},\ and\ \citenamefont
  {Sorella}}]{Capriotti_Trumper_Adolfo_Sorella_PRL_1999}%
  \BibitemOpen
  \bibfield  {author} {\bibinfo {author} {\bibfnamefont {L.}~\bibnamefont
  {Capriotti}}, \bibinfo {author} {\bibfnamefont {A.~E.}\ \bibnamefont
  {Trumper}},\ and\ \bibinfo {author} {\bibfnamefont {S.}~\bibnamefont
  {Sorella}},\ }\bibfield  {title} {\bibinfo {title} {Long-range {N\'eel} order
  in the triangular {Heisenberg} model},\ }\href
  {https://doi.org/10.1103/PhysRevLett.82.3899} {\bibfield  {journal} {\bibinfo
   {journal} {Phys. Rev. Lett.}\ }\textbf {\bibinfo {volume} {82}},\ \bibinfo
  {pages} {3899} (\bibinfo {year} {1999})}\BibitemShut {NoStop}%
\bibitem [{\citenamefont {White}\ and\ \citenamefont
  {Chernyshev}(2007)}]{White_Chernyshev_PRL_2007}%
  \BibitemOpen
  \bibfield  {author} {\bibinfo {author} {\bibfnamefont {S.~R.}\ \bibnamefont
  {White}}\ and\ \bibinfo {author} {\bibfnamefont {A.~L.}\ \bibnamefont
  {Chernyshev}},\ }\bibfield  {title} {\bibinfo {title} {{Ne\'el} order in
  square and triangular lattice {Heisenberg} models},\ }\href
  {https://doi.org/10.1103/PhysRevLett.99.127004} {\bibfield  {journal}
  {\bibinfo  {journal} {Phys. Rev. Lett.}\ }\textbf {\bibinfo {volume} {99}},\
  \bibinfo {pages} {127004} (\bibinfo {year} {2007})}\BibitemShut {NoStop}%
\bibitem [{\citenamefont {Kivelson}\ \emph {et~al.}(1987)\citenamefont
  {Kivelson}, \citenamefont {Rokhsar},\ and\ \citenamefont
  {Sethna}}]{Kivelson_Rokhsar_Sethna_PRB_1987}%
  \BibitemOpen
  \bibfield  {author} {\bibinfo {author} {\bibfnamefont {S.~A.}\ \bibnamefont
  {Kivelson}}, \bibinfo {author} {\bibfnamefont {D.~S.}\ \bibnamefont
  {Rokhsar}},\ and\ \bibinfo {author} {\bibfnamefont {J.~P.}\ \bibnamefont
  {Sethna}},\ }\bibfield  {title} {\bibinfo {title} {Topology of the resonating
  valence-bond state: Solitons and high-${T}_{c}$ superconductivity},\ }\href
  {https://doi.org/10.1103/PhysRevB.35.8865} {\bibfield  {journal} {\bibinfo
  {journal} {Phys. Rev. B}\ }\textbf {\bibinfo {volume} {35}},\ \bibinfo
  {pages} {8865} (\bibinfo {year} {1987})}\BibitemShut {NoStop}%
\bibitem [{\citenamefont {Rokhsar}\ and\ \citenamefont
  {Kivelson}(1988)}]{Rokhsar_Kivelson_PRL_1988}%
  \BibitemOpen
  \bibfield  {author} {\bibinfo {author} {\bibfnamefont {D.~S.}\ \bibnamefont
  {Rokhsar}}\ and\ \bibinfo {author} {\bibfnamefont {S.~A.}\ \bibnamefont
  {Kivelson}},\ }\bibfield  {title} {\bibinfo {title} {Superconductivity and
  the quantum hard-core dimer gas},\ }\href
  {https://doi.org/10.1103/PhysRevLett.61.2376} {\bibfield  {journal} {\bibinfo
   {journal} {Phys. Rev. Lett.}\ }\textbf {\bibinfo {volume} {61}},\ \bibinfo
  {pages} {2376} (\bibinfo {year} {1988})}\BibitemShut {NoStop}%
\bibitem [{\citenamefont {Moessner}\ and\ \citenamefont
  {Sondhi}(2001)}]{Moessner_Sondhi_PRL_2001}%
  \BibitemOpen
  \bibfield  {author} {\bibinfo {author} {\bibfnamefont {R.}~\bibnamefont
  {Moessner}}\ and\ \bibinfo {author} {\bibfnamefont {S.~L.}\ \bibnamefont
  {Sondhi}},\ }\bibfield  {title} {\bibinfo {title} {Resonating valence bond
  phase in the triangular lattice quantum dimer model},\ }\href
  {https://doi.org/10.1103/PhysRevLett.86.1881} {\bibfield  {journal} {\bibinfo
   {journal} {Phys. Rev. Lett.}\ }\textbf {\bibinfo {volume} {86}},\ \bibinfo
  {pages} {1881} (\bibinfo {year} {2001})}\BibitemShut {NoStop}%
\bibitem [{\citenamefont {Kalmeyer}\ and\ \citenamefont
  {Laughlin}(1987)}]{Kalmeyer_Laughlin_PRL_1987}%
  \BibitemOpen
  \bibfield  {author} {\bibinfo {author} {\bibfnamefont {V.}~\bibnamefont
  {Kalmeyer}}\ and\ \bibinfo {author} {\bibfnamefont {R.~B.}\ \bibnamefont
  {Laughlin}},\ }\bibfield  {title} {\bibinfo {title} {Equivalence of the
  resonating-valence-bond and fractional quantum {H}all states},\ }\href
  {https://doi.org/10.1103/PhysRevLett.59.2095} {\bibfield  {journal} {\bibinfo
   {journal} {Phys. Rev. Lett.}\ }\textbf {\bibinfo {volume} {59}},\ \bibinfo
  {pages} {2095} (\bibinfo {year} {1987})}\BibitemShut {NoStop}%
\bibitem [{\citenamefont {Affleck}\ and\ \citenamefont
  {Marston}(1988)}]{Affleck_Marston_PRB_1988}%
  \BibitemOpen
  \bibfield  {author} {\bibinfo {author} {\bibfnamefont {I.}~\bibnamefont
  {Affleck}}\ and\ \bibinfo {author} {\bibfnamefont {J.~B.}\ \bibnamefont
  {Marston}},\ }\bibfield  {title} {\bibinfo {title} {Large-n limit of the
  {H}eisenberg-{H}ubbard model: Implications for high-${T}_{c}$
  superconductors},\ }\href {https://doi.org/10.1103/PhysRevB.37.3774}
  {\bibfield  {journal} {\bibinfo  {journal} {Phys. Rev. B}\ }\textbf {\bibinfo
  {volume} {37}},\ \bibinfo {pages} {3774} (\bibinfo {year}
  {1988})}\BibitemShut {NoStop}%
\bibitem [{\citenamefont {Read}\ and\ \citenamefont
  {Sachdev}(1989)}]{Read_Sachdev_NuPhysB_1989}%
  \BibitemOpen
  \bibfield  {author} {\bibinfo {author} {\bibfnamefont {N.}~\bibnamefont
  {Read}}\ and\ \bibinfo {author} {\bibfnamefont {S.}~\bibnamefont {Sachdev}},\
  }\bibfield  {title} {\bibinfo {title} {Some features of the phase diagram of
  the square lattice {SU(N)} antiferromagnet},\ }\href
  {https://doi.org/https://doi.org/10.1016/0550-3213(89)90061-8} {\bibfield
  {journal} {\bibinfo  {journal} {Nuclear Physics B}\ }\textbf {\bibinfo
  {volume} {316}},\ \bibinfo {pages} {609} (\bibinfo {year}
  {1989})}\BibitemShut {NoStop}%
\bibitem [{\citenamefont {Read}\ and\ \citenamefont
  {Sachdev}(1991)}]{Read_Sachdev_PRL_1991}%
  \BibitemOpen
  \bibfield  {author} {\bibinfo {author} {\bibfnamefont {N.}~\bibnamefont
  {Read}}\ and\ \bibinfo {author} {\bibfnamefont {S.}~\bibnamefont {Sachdev}},\
  }\bibfield  {title} {\bibinfo {title} {Large-{N} expansion for frustrated
  quantum antiferromagnets},\ }\href
  {https://doi.org/10.1103/PhysRevLett.66.1773} {\bibfield  {journal} {\bibinfo
   {journal} {Phys. Rev. Lett.}\ }\textbf {\bibinfo {volume} {66}},\ \bibinfo
  {pages} {1773} (\bibinfo {year} {1991})}\BibitemShut {NoStop}%
\bibitem [{\citenamefont {Sachdev}(1992)}]{Sachdev_PRB_1992}%
  \BibitemOpen
  \bibfield  {author} {\bibinfo {author} {\bibfnamefont {S.}~\bibnamefont
  {Sachdev}},\ }\bibfield  {title} {\bibinfo {title} {Kagome and
  triangular-lattice {H}eisenberg antiferromagnets: Ordering from quantum
  fluctuations and quantum-disordered ground states with unconfined bosonic
  spinons},\ }\href {https://doi.org/10.1103/PhysRevB.45.12377} {\bibfield
  {journal} {\bibinfo  {journal} {Phys. Rev. B}\ }\textbf {\bibinfo {volume}
  {45}},\ \bibinfo {pages} {12377} (\bibinfo {year} {1992})}\BibitemShut
  {NoStop}%
\bibitem [{\citenamefont {Misguich}\ \emph {et~al.}(1999)\citenamefont
  {Misguich}, \citenamefont {Lhuillier}, \citenamefont {Bernu},\ and\
  \citenamefont {Waldtmann}}]{Misguich_Lhuillier_Bernu_Waldtmann_PRB_1999}%
  \BibitemOpen
  \bibfield  {author} {\bibinfo {author} {\bibfnamefont {G.}~\bibnamefont
  {Misguich}}, \bibinfo {author} {\bibfnamefont {C.}~\bibnamefont {Lhuillier}},
  \bibinfo {author} {\bibfnamefont {B.}~\bibnamefont {Bernu}},\ and\ \bibinfo
  {author} {\bibfnamefont {C.}~\bibnamefont {Waldtmann}},\ }\bibfield  {title}
  {\bibinfo {title} {Spin-liquid phase of the multiple-spin exchange
  {H}amiltonian on the triangular lattice},\ }\href
  {https://doi.org/10.1103/PhysRevB.60.1064} {\bibfield  {journal} {\bibinfo
  {journal} {Phys. Rev. B}\ }\textbf {\bibinfo {volume} {60}},\ \bibinfo
  {pages} {1064} (\bibinfo {year} {1999})}\BibitemShut {NoStop}%
\bibitem [{\citenamefont {Wen}(1991)}]{Wen_PRB_1991}%
  \BibitemOpen
  \bibfield  {author} {\bibinfo {author} {\bibfnamefont {X.~G.}\ \bibnamefont
  {Wen}},\ }\bibfield  {title} {\bibinfo {title} {Mean-field theory of
  spin-liquid states with finite energy gap and topological orders},\ }\href
  {https://doi.org/10.1103/PhysRevB.44.2664} {\bibfield  {journal} {\bibinfo
  {journal} {Phys. Rev. B}\ }\textbf {\bibinfo {volume} {44}},\ \bibinfo
  {pages} {2664} (\bibinfo {year} {1991})}\BibitemShut {NoStop}%
\bibitem [{\citenamefont {Wen}(2007)}]{Wen_QFTbook_2007}%
  \BibitemOpen
  \bibfield  {author} {\bibinfo {author} {\bibfnamefont {X.}~\bibnamefont
  {Wen}},\ }\href {https://doi.org/10.1093/acprof:oso/9780199227259.001.0001}
  {\emph {\bibinfo {title} {{Quantum Field Theory of Many-Body Systems: From
  the Origin of Sound to an Origin of Light and Electrons}}}}\ (\bibinfo
  {publisher} {Oxford University Press},\ \bibinfo {year} {2007})\BibitemShut
  {NoStop}%
\bibitem [{\citenamefont {Wen}(2017)}]{Wen_RevModPhys_2017}%
  \BibitemOpen
  \bibfield  {author} {\bibinfo {author} {\bibfnamefont {X.}~\bibnamefont
  {Wen}},\ }\bibfield  {title} {\bibinfo {title} {Colloquium: Zoo of
  quantum-topological phases of matter},\ }\href
  {https://doi.org/10.1103/RevModPhys.89.041004} {\bibfield  {journal}
  {\bibinfo  {journal} {Rev. Mod. Phys.}\ }\textbf {\bibinfo {volume} {89}},\
  \bibinfo {pages} {041004} (\bibinfo {year} {2017})}\BibitemShut {NoStop}%
\bibitem [{\citenamefont {Moessner}\ and\ \citenamefont
  {Moore}(2021)}]{Moessner_Moore_TPMbook_2021}%
  \BibitemOpen
  \bibfield  {author} {\bibinfo {author} {\bibfnamefont {R.}~\bibnamefont
  {Moessner}}\ and\ \bibinfo {author} {\bibfnamefont {J.~E.}\ \bibnamefont
  {Moore}},\ }\href {https://doi.org/https://doi.org/10.1017/9781316226308}
  {\emph {\bibinfo {title} {{Topological Phases of Matter}}}}\ (\bibinfo
  {publisher} {Cambridge University Press},\ \bibinfo {year}
  {2021})\BibitemShut {NoStop}%
\bibitem [{\citenamefont {Sachdev}(2023)}]{Sachdev_QPMbook_2023}%
  \BibitemOpen
  \bibfield  {author} {\bibinfo {author} {\bibfnamefont {S.}~\bibnamefont
  {Sachdev}},\ }\href {https://doi.org/https://doi.org/10.1017/9781009212717}
  {\emph {\bibinfo {title} {{Quantum Phases of Matter}}}}\ (\bibinfo
  {publisher} {Cambridge University Press},\ \bibinfo {year}
  {2023})\BibitemShut {NoStop}%
\bibitem [{\citenamefont {Norman}(2016)}]{Norman_RevModPhys_2016}%
  \BibitemOpen
  \bibfield  {author} {\bibinfo {author} {\bibfnamefont {M.~R.}\ \bibnamefont
  {Norman}},\ }\bibfield  {title} {\bibinfo {title} {Colloquium:
  Herbertsmithite and the search for the quantum spin liquid},\ }\href
  {https://doi.org/10.1103/RevModPhys.88.041002} {\bibfield  {journal}
  {\bibinfo  {journal} {Rev. Mod. Phys.}\ }\textbf {\bibinfo {volume} {88}},\
  \bibinfo {pages} {041002} (\bibinfo {year} {2016})}\BibitemShut {NoStop}%
\bibitem [{\citenamefont {Broholm}\ \emph {et~al.}(2020)\citenamefont
  {Broholm}, \citenamefont {Cava}, \citenamefont {Kivelson}, \citenamefont
  {Nocera}, \citenamefont {Norman},\ and\ \citenamefont
  {Senthil}}]{Broholm_Cava_Kivelson_Nocera_Norman_Senthil_Science_2020}%
  \BibitemOpen
  \bibfield  {author} {\bibinfo {author} {\bibfnamefont {C.}~\bibnamefont
  {Broholm}}, \bibinfo {author} {\bibfnamefont {R.~J.}\ \bibnamefont {Cava}},
  \bibinfo {author} {\bibfnamefont {S.~A.}\ \bibnamefont {Kivelson}}, \bibinfo
  {author} {\bibfnamefont {D.~G.}\ \bibnamefont {Nocera}}, \bibinfo {author}
  {\bibfnamefont {M.~R.}\ \bibnamefont {Norman}},\ and\ \bibinfo {author}
  {\bibfnamefont {T.}~\bibnamefont {Senthil}},\ }\bibfield  {title} {\bibinfo
  {title} {Quantum spin liquids},\ }\href
  {https://doi.org/10.1126/science.aay0668} {\bibfield  {journal} {\bibinfo
  {journal} {Science}\ }\textbf {\bibinfo {volume} {367}},\ \bibinfo {pages}
  {eaay0668} (\bibinfo {year} {2020})}\BibitemShut {NoStop}%
\bibitem [{\citenamefont {Clark}\ and\ \citenamefont
  {Abdeldaim}(2021)}]{Clark_Abdeldaim_Annual_Review_of_Material_Research_2021}%
  \BibitemOpen
  \bibfield  {author} {\bibinfo {author} {\bibfnamefont {L.}~\bibnamefont
  {Clark}}\ and\ \bibinfo {author} {\bibfnamefont {A.~H.}\ \bibnamefont
  {Abdeldaim}},\ }\bibfield  {title} {\bibinfo {title} {Quantum spin liquids
  from a materials perspective},\ }\href
  {https://doi.org/10.1146/annurev-matsci-080819-011453} {\bibfield  {journal}
  {\bibinfo  {journal} {Annual Review of Materials Research}\ }\textbf
  {\bibinfo {volume} {51}},\ \bibinfo {pages} {495} (\bibinfo {year}
  {2021})}\BibitemShut {NoStop}%
\bibitem [{\citenamefont {Lee}\ \emph {et~al.}(2007)\citenamefont {Lee},
  \citenamefont {Kikuchi}, \citenamefont {Qiu}, \citenamefont {Lake},
  \citenamefont {Huang}, \citenamefont {Habicht},\ and\ \citenamefont
  {Kiefer}}]{Lee_nature_2007}%
  \BibitemOpen
  \bibfield  {author} {\bibinfo {author} {\bibfnamefont {S.-H.}\ \bibnamefont
  {Lee}}, \bibinfo {author} {\bibfnamefont {H.}~\bibnamefont {Kikuchi}},
  \bibinfo {author} {\bibfnamefont {Y.}~\bibnamefont {Qiu}}, \bibinfo {author}
  {\bibfnamefont {B.}~\bibnamefont {Lake}}, \bibinfo {author} {\bibfnamefont
  {Q.}~\bibnamefont {Huang}}, \bibinfo {author} {\bibfnamefont
  {K.}~\bibnamefont {Habicht}},\ and\ \bibinfo {author} {\bibfnamefont
  {K.}~\bibnamefont {Kiefer}},\ }\bibfield  {title} {\bibinfo {title}
  {Quantum-spin-liquid states in the two-dimensional kagome antiferromagnets
  {${\mathrm{Zn}}_{x}{\mathrm{Cu}}_{4\ensuremath{-}x}{\mathrm{(OD)}}_{6}{\mathrm{Cl}}_{2}$}},\
  }\href {https://doi.org/10.1038/nmat1986} {\bibfield  {journal} {\bibinfo
  {journal} {Nature Materials}\ }\textbf {\bibinfo {volume} {6}},\ \bibinfo
  {pages} {853} (\bibinfo {year} {2007})}\BibitemShut {NoStop}%
\bibitem [{\citenamefont {de~Vries}\ \emph {et~al.}(2008)\citenamefont
  {de~Vries}, \citenamefont {Kamenev}, \citenamefont {Kockelmann},
  \citenamefont {Sanchez-Benitez},\ and\ \citenamefont
  {Harrison}}]{Vries_Kamenev_Kockelmann_Benitez_Harrison_PRL_2008}%
  \BibitemOpen
  \bibfield  {author} {\bibinfo {author} {\bibfnamefont {M.~A.}\ \bibnamefont
  {de~Vries}}, \bibinfo {author} {\bibfnamefont {K.~V.}\ \bibnamefont
  {Kamenev}}, \bibinfo {author} {\bibfnamefont {W.~A.}\ \bibnamefont
  {Kockelmann}}, \bibinfo {author} {\bibfnamefont {J.}~\bibnamefont
  {Sanchez-Benitez}},\ and\ \bibinfo {author} {\bibfnamefont {A.}~\bibnamefont
  {Harrison}},\ }\bibfield  {title} {\bibinfo {title} {Magnetic ground state of
  an experimental {$S=1/2$} kagome antiferromagnet},\ }\href
  {https://doi.org/10.1103/PhysRevLett.100.157205} {\bibfield  {journal}
  {\bibinfo  {journal} {Phys. Rev. Lett.}\ }\textbf {\bibinfo {volume} {100}},\
  \bibinfo {pages} {157205} (\bibinfo {year} {2008})}\BibitemShut {NoStop}%
\bibitem [{\citenamefont {H\"au\ss{}ler}\ \emph {et~al.}(2022)\citenamefont
  {H\"au\ss{}ler}, \citenamefont {Sichelschmidt}, \citenamefont {Baenitz},
  \citenamefont {Andrade}, \citenamefont {Vojta},\ and\ \citenamefont
  {Doert}}]{Ellen_Vojta_Doert_PRM_2022}%
  \BibitemOpen
  \bibfield  {author} {\bibinfo {author} {\bibfnamefont {E.}~\bibnamefont
  {H\"au\ss{}ler}}, \bibinfo {author} {\bibfnamefont {J.}~\bibnamefont
  {Sichelschmidt}}, \bibinfo {author} {\bibfnamefont {M.}~\bibnamefont
  {Baenitz}}, \bibinfo {author} {\bibfnamefont {E.~C.}\ \bibnamefont
  {Andrade}}, \bibinfo {author} {\bibfnamefont {M.}~\bibnamefont {Vojta}},\
  and\ \bibinfo {author} {\bibfnamefont {T.}~\bibnamefont {Doert}},\ }\bibfield
   {title} {\bibinfo {title} {Diluting a triangular-lattice spin liquid:
  Synthesis and characterization of
  {${\mathrm{NaYb}}_{1\ensuremath{-}x}{\mathrm{Lu}}_{x}{\mathrm{S}}_{2}$}
  single crystals},\ }\href {https://doi.org/10.1103/PhysRevMaterials.6.046201}
  {\bibfield  {journal} {\bibinfo  {journal} {Phys. Rev. Mater.}\ }\textbf
  {\bibinfo {volume} {6}},\ \bibinfo {pages} {046201} (\bibinfo {year}
  {2022})}\BibitemShut {NoStop}%
\bibitem [{\citenamefont {Paddison}\ \emph {et~al.}(2017)\citenamefont
  {Paddison}, \citenamefont {Daum}, \citenamefont {Dun}, \citenamefont
  {Ehlers}, \citenamefont {Liu}, \citenamefont {Stone}, \citenamefont {Zhou},\
  and\ \citenamefont
  {Mourigal}}]{Paddison_Daum_Dun_Ehlers_Liu_Stone_Zhou_Mourigal_Nature_Physics_2017}%
  \BibitemOpen
  \bibfield  {author} {\bibinfo {author} {\bibfnamefont {J.~A.~M.}\
  \bibnamefont {Paddison}}, \bibinfo {author} {\bibfnamefont {M.}~\bibnamefont
  {Daum}}, \bibinfo {author} {\bibfnamefont {Z.}~\bibnamefont {Dun}}, \bibinfo
  {author} {\bibfnamefont {G.}~\bibnamefont {Ehlers}}, \bibinfo {author}
  {\bibfnamefont {Y.}~\bibnamefont {Liu}}, \bibinfo {author} {\bibfnamefont
  {M.~B.}\ \bibnamefont {Stone}}, \bibinfo {author} {\bibfnamefont
  {H.}~\bibnamefont {Zhou}},\ and\ \bibinfo {author} {\bibfnamefont
  {M.}~\bibnamefont {Mourigal}},\ }\bibfield  {title} {\bibinfo {title}
  {Continuous excitations of the triangular-lattice quantum spin liquid
  {${\mathrm{YbMgGaO}}_{4}$}},\ }\href {https://doi.org/10.1038/nphys3971}
  {\bibfield  {journal} {\bibinfo  {journal} {Nature Physics}\ }\textbf
  {\bibinfo {volume} {13}},\ \bibinfo {pages} {117} (\bibinfo {year}
  {2017})}\BibitemShut {NoStop}%
\bibitem [{\citenamefont {Kimchi}\ \emph {et~al.}(2018)\citenamefont {Kimchi},
  \citenamefont {Nahum},\ and\ \citenamefont
  {Senthil}}]{Kimchi_Nahum_Senthil_PRX_2018}%
  \BibitemOpen
  \bibfield  {author} {\bibinfo {author} {\bibfnamefont {I.}~\bibnamefont
  {Kimchi}}, \bibinfo {author} {\bibfnamefont {A.}~\bibnamefont {Nahum}},\ and\
  \bibinfo {author} {\bibfnamefont {T.}~\bibnamefont {Senthil}},\ }\bibfield
  {title} {\bibinfo {title} {Valence bonds in random quantum magnets: Theory
  and application to {${\mathrm{YbMgGaO}}_{4}$}},\ }\href
  {https://doi.org/10.1103/PhysRevX.8.031028} {\bibfield  {journal} {\bibinfo
  {journal} {Phys. Rev. X}\ }\textbf {\bibinfo {volume} {8}},\ \bibinfo {pages}
  {031028} (\bibinfo {year} {2018})}\BibitemShut {NoStop}%
\bibitem [{\citenamefont {Sandvik}\ \emph {et~al.}(1997)\citenamefont
  {Sandvik}, \citenamefont {Dagotto},\ and\ \citenamefont
  {Scalapino}}]{Sandvik_Dagotto_Scalapino_PRB_1997}%
  \BibitemOpen
  \bibfield  {author} {\bibinfo {author} {\bibfnamefont {A.~W.}\ \bibnamefont
  {Sandvik}}, \bibinfo {author} {\bibfnamefont {E.}~\bibnamefont {Dagotto}},\
  and\ \bibinfo {author} {\bibfnamefont {D.~J.}\ \bibnamefont {Scalapino}},\
  }\bibfield  {title} {\bibinfo {title} {Nonmagnetic impurities in spin-gapped
  and gapless {H}eisenberg antiferromagnets},\ }\href
  {https://doi.org/10.1103/PhysRevB.56.11701} {\bibfield  {journal} {\bibinfo
  {journal} {Phys. Rev. B}\ }\textbf {\bibinfo {volume} {56}},\ \bibinfo
  {pages} {11701} (\bibinfo {year} {1997})}\BibitemShut {NoStop}%
\bibitem [{\citenamefont {Sachdev}\ \emph {et~al.}(1999)\citenamefont
  {Sachdev}, \citenamefont {Buragohain},\ and\ \citenamefont
  {Vojta}}]{Sachdev_Buragohain_Vojta_Science_1999}%
  \BibitemOpen
  \bibfield  {author} {\bibinfo {author} {\bibfnamefont {S.}~\bibnamefont
  {Sachdev}}, \bibinfo {author} {\bibfnamefont {C.}~\bibnamefont
  {Buragohain}},\ and\ \bibinfo {author} {\bibfnamefont {M.}~\bibnamefont
  {Vojta}},\ }\bibfield  {title} {\bibinfo {title} {Quantum impurity in a
  nearly critical two-dimensional antiferromagnet},\ }\href
  {https://doi.org/10.1126/science.286.5449.2479} {\bibfield  {journal}
  {\bibinfo  {journal} {Science}\ }\textbf {\bibinfo {volume} {286}},\ \bibinfo
  {pages} {2479} (\bibinfo {year} {1999})}\BibitemShut {NoStop}%
\bibitem [{\citenamefont {Gregor}\ and\ \citenamefont
  {Motrunich}(2009)}]{Gregor_Motrunich_PRB_2009}%
  \BibitemOpen
  \bibfield  {author} {\bibinfo {author} {\bibfnamefont {K.}~\bibnamefont
  {Gregor}}\ and\ \bibinfo {author} {\bibfnamefont {O.~I.}\ \bibnamefont
  {Motrunich}},\ }\bibfield  {title} {\bibinfo {title} {Nonmagnetic impurities
  in a {$\rm S=\frac{1}{2}$} frustrated triangular antiferromagnet: Broadening
  of $^{13}\text{C}$ {NMR} lines in
  $\ensuremath{\kappa}\text{\ensuremath{-}}{(\text{ET})}_{2}{{\text{Cu}}}_{2}{(\text{CN})}_{3}$},\
  }\href {https://doi.org/10.1103/PhysRevB.79.024421} {\bibfield  {journal}
  {\bibinfo  {journal} {Phys. Rev. B}\ }\textbf {\bibinfo {volume} {79}},\
  \bibinfo {pages} {024421} (\bibinfo {year} {2009})}\BibitemShut {NoStop}%
\bibitem [{\citenamefont {Wang}\ and\ \citenamefont
  {Sandvik}(2010)}]{Wang_Sandvik_PRB_2010}%
  \BibitemOpen
  \bibfield  {author} {\bibinfo {author} {\bibfnamefont {L.}~\bibnamefont
  {Wang}}\ and\ \bibinfo {author} {\bibfnamefont {A.~W.}\ \bibnamefont
  {Sandvik}},\ }\bibfield  {title} {\bibinfo {title} {Low-energy excitations of
  two-dimensional diluted {H}eisenberg quantum antiferromagnets},\ }\href
  {https://doi.org/10.1103/PhysRevB.81.054417} {\bibfield  {journal} {\bibinfo
  {journal} {Phys. Rev. B}\ }\textbf {\bibinfo {volume} {81}},\ \bibinfo
  {pages} {054417} (\bibinfo {year} {2010})}\BibitemShut {NoStop}%
\bibitem [{\citenamefont {Ghosh}\ \emph {et~al.}(2015)\citenamefont {Ghosh},
  \citenamefont {Changlani},\ and\ \citenamefont
  {Henley}}]{Ghosh_Changlani_Henley_PRB_2015}%
  \BibitemOpen
  \bibfield  {author} {\bibinfo {author} {\bibfnamefont {S.}~\bibnamefont
  {Ghosh}}, \bibinfo {author} {\bibfnamefont {H.~J.}\ \bibnamefont
  {Changlani}},\ and\ \bibinfo {author} {\bibfnamefont {C.~L.}\ \bibnamefont
  {Henley}},\ }\bibfield  {title} {\bibinfo {title} {Schwinger boson mean field
  perspective on emergent spins in diluted {H}eisenberg antiferromagnets},\
  }\href {https://doi.org/10.1103/PhysRevB.92.064401} {\bibfield  {journal}
  {\bibinfo  {journal} {Phys. Rev. B}\ }\textbf {\bibinfo {volume} {92}},\
  \bibinfo {pages} {064401} (\bibinfo {year} {2015})}\BibitemShut {NoStop}%
\bibitem [{\citenamefont {H\"oglund}\ \emph {et~al.}(2007)\citenamefont
  {H\"oglund}, \citenamefont {Sandvik},\ and\ \citenamefont
  {Sachdev}}]{Hoglund_Sandvik_Sachdev_PRL_2007}%
  \BibitemOpen
  \bibfield  {author} {\bibinfo {author} {\bibfnamefont {K.~H.}\ \bibnamefont
  {H\"oglund}}, \bibinfo {author} {\bibfnamefont {A.~W.}\ \bibnamefont
  {Sandvik}},\ and\ \bibinfo {author} {\bibfnamefont {S.}~\bibnamefont
  {Sachdev}},\ }\bibfield  {title} {\bibinfo {title} {Impurity induced spin
  texture in quantum critical {2D} antiferromagnets},\ }\href
  {https://doi.org/10.1103/PhysRevLett.98.087203} {\bibfield  {journal}
  {\bibinfo  {journal} {Phys. Rev. Lett.}\ }\textbf {\bibinfo {volume} {98}},\
  \bibinfo {pages} {087203} (\bibinfo {year} {2007})}\BibitemShut {NoStop}%
\bibitem [{\citenamefont {Kaul}\ \emph {et~al.}(2008)\citenamefont {Kaul},
  \citenamefont {Melko}, \citenamefont {Metlitski},\ and\ \citenamefont
  {Sachdev}}]{Kaul_Melko_Metlitski_Sachdev_PRL_2008}%
  \BibitemOpen
  \bibfield  {author} {\bibinfo {author} {\bibfnamefont {R.~K.}\ \bibnamefont
  {Kaul}}, \bibinfo {author} {\bibfnamefont {R.~G.}\ \bibnamefont {Melko}},
  \bibinfo {author} {\bibfnamefont {M.~A.}\ \bibnamefont {Metlitski}},\ and\
  \bibinfo {author} {\bibfnamefont {S.}~\bibnamefont {Sachdev}},\ }\bibfield
  {title} {\bibinfo {title} {Imaging bond order near nonmagnetic impurities in
  square-lattice antiferromagnets},\ }\href
  {https://doi.org/10.1103/PhysRevLett.101.187206} {\bibfield  {journal}
  {\bibinfo  {journal} {Phys. Rev. Lett.}\ }\textbf {\bibinfo {volume} {101}},\
  \bibinfo {pages} {187206} (\bibinfo {year} {2008})}\BibitemShut {NoStop}%
\bibitem [{\citenamefont {Banerjee}\ \emph {et~al.}(2010)\citenamefont
  {Banerjee}, \citenamefont {Damle},\ and\ \citenamefont
  {Alet}}]{Banerjee_Damle_Alet_PRB_2010}%
  \BibitemOpen
  \bibfield  {author} {\bibinfo {author} {\bibfnamefont {A.}~\bibnamefont
  {Banerjee}}, \bibinfo {author} {\bibfnamefont {K.}~\bibnamefont {Damle}},\
  and\ \bibinfo {author} {\bibfnamefont {F.}~\bibnamefont {Alet}},\ }\bibfield
  {title} {\bibinfo {title} {Impurity spin texture at a deconfined quantum
  critical point},\ }\href {https://doi.org/10.1103/PhysRevB.82.155139}
  {\bibfield  {journal} {\bibinfo  {journal} {Phys. Rev. B}\ }\textbf {\bibinfo
  {volume} {82}},\ \bibinfo {pages} {155139} (\bibinfo {year}
  {2010})}\BibitemShut {NoStop}%
\bibitem [{\citenamefont {Sanyal}\ \emph {et~al.}(2011)\citenamefont {Sanyal},
  \citenamefont {Banerjee},\ and\ \citenamefont
  {Damle}}]{Sanyal_Banerjee_Damle_PRB_2011}%
  \BibitemOpen
  \bibfield  {author} {\bibinfo {author} {\bibfnamefont {S.}~\bibnamefont
  {Sanyal}}, \bibinfo {author} {\bibfnamefont {A.}~\bibnamefont {Banerjee}},\
  and\ \bibinfo {author} {\bibfnamefont {K.}~\bibnamefont {Damle}},\ }\bibfield
   {title} {\bibinfo {title} {Vacancy-induced spin texture in a one-dimensional
  {$\rm S=\frac{1}{2}$} {H}eisenberg antiferromagnet},\ }\href
  {https://doi.org/10.1103/PhysRevB.84.235129} {\bibfield  {journal} {\bibinfo
  {journal} {Phys. Rev. B}\ }\textbf {\bibinfo {volume} {84}},\ \bibinfo
  {pages} {235129} (\bibinfo {year} {2011})}\BibitemShut {NoStop}%
\bibitem [{\citenamefont {Banerjee}\ \emph
  {et~al.}(2011{\natexlab{a}})\citenamefont {Banerjee}, \citenamefont {Damle},\
  and\ \citenamefont {Alet}}]{Banerjee_Damle_Alet_PRB_2011}%
  \BibitemOpen
  \bibfield  {author} {\bibinfo {author} {\bibfnamefont {A.}~\bibnamefont
  {Banerjee}}, \bibinfo {author} {\bibfnamefont {K.}~\bibnamefont {Damle}},\
  and\ \bibinfo {author} {\bibfnamefont {F.}~\bibnamefont {Alet}},\ }\bibfield
  {title} {\bibinfo {title} {{Impurity spin texture at the critical point
  between N\'eel-ordered and valence-bond-solid states in two-dimensional SU(3)
  quantum antiferromagnets}},\ }\href
  {https://doi.org/10.1103/PhysRevB.83.235111} {\bibfield  {journal} {\bibinfo
  {journal} {Phys. Rev. B}\ }\textbf {\bibinfo {volume} {83}},\ \bibinfo
  {pages} {235111} (\bibinfo {year} {2011}{\natexlab{a}})}\BibitemShut
  {NoStop}%
\bibitem [{\citenamefont {Aldred}\ \emph {et~al.}(2007)\citenamefont {Aldred},
  \citenamefont {Anstee},\ and\ \citenamefont
  {Locke}}]{Aldred_Anstee_Locke_Discrete_Mathematics_2007}%
  \BibitemOpen
  \bibfield  {author} {\bibinfo {author} {\bibfnamefont {R.}~\bibnamefont
  {Aldred}}, \bibinfo {author} {\bibfnamefont {R.}~\bibnamefont {Anstee}},\
  and\ \bibinfo {author} {\bibfnamefont {S.}~\bibnamefont {Locke}},\ }\bibfield
   {title} {\bibinfo {title} {Perfect matchings after vertex deletions},\
  }\href {https://doi.org/https://doi.org/10.1016/j.disc.2007.03.017}
  {\bibfield  {journal} {\bibinfo  {journal} {Discrete Mathematics}\ }\textbf
  {\bibinfo {volume} {307}},\ \bibinfo {pages} {3048} (\bibinfo {year}
  {2007})}\BibitemShut {NoStop}%
\bibitem [{\citenamefont {Anstee}\ \emph {et~al.}(2011)\citenamefont {Anstee},
  \citenamefont {Blackman},\ and\ \citenamefont
  {Yang}}]{Anstee_Blackman_Yang_Discrete_Mathematics_2011}%
  \BibitemOpen
  \bibfield  {author} {\bibinfo {author} {\bibfnamefont {R.~P.}\ \bibnamefont
  {Anstee}}, \bibinfo {author} {\bibfnamefont {J.}~\bibnamefont {Blackman}},\
  and\ \bibinfo {author} {\bibfnamefont {H.}~\bibnamefont {Yang}},\ }\bibfield
  {title} {\bibinfo {title} {Perfect matchings in grid graphs after vertex
  deletions},\ }\href {https://doi.org/10.1137/090776706} {\bibfield  {journal}
  {\bibinfo  {journal} {SIAM Journal on Discrete Mathematics}\ }\textbf
  {\bibinfo {volume} {25}},\ \bibinfo {pages} {1754} (\bibinfo {year}
  {2011})}\BibitemShut {NoStop}%
\bibitem [{\citenamefont {Anstee}\ \emph {et~al.}(2006)\citenamefont {Anstee},
  \citenamefont {Blackman},\ and\ \citenamefont {Yang}}]{Tseng_Anstee_2006}%
  \BibitemOpen
  \bibfield  {author} {\bibinfo {author} {\bibfnamefont {R.~P.}\ \bibnamefont
  {Anstee}}, \bibinfo {author} {\bibfnamefont {J.}~\bibnamefont {Blackman}},\
  and\ \bibinfo {author} {\bibfnamefont {H.}~\bibnamefont {Yang}},\ }\bibfield
  {title} {\bibinfo {title} {Perfect matching after vertex deletions on the
  grid graph and triangular graph (unpublished)},\ }\href@noop {} {\bibfield
  {journal} {\bibinfo  {journal} {preprint}\ } (\bibinfo {year}
  {2006})}\BibitemShut {NoStop}%
\bibitem [{\citenamefont {Bhola}\ \emph {et~al.}(2022)\citenamefont {Bhola},
  \citenamefont {Biswas}, \citenamefont {Islam},\ and\ \citenamefont
  {Damle}}]{Bhola_Biswas_Islam_Damle_PRX_2022}%
  \BibitemOpen
  \bibfield  {author} {\bibinfo {author} {\bibfnamefont {R.}~\bibnamefont
  {Bhola}}, \bibinfo {author} {\bibfnamefont {S.}~\bibnamefont {Biswas}},
  \bibinfo {author} {\bibfnamefont {M.~M.}\ \bibnamefont {Islam}},\ and\
  \bibinfo {author} {\bibfnamefont {K.}~\bibnamefont {Damle}},\ }\bibfield
  {title} {\bibinfo {title} {Dulmage-{M}endelsohn percolation: Geometry of
  maximally packed dimer models and topologically protected zero modes on
  site-diluted bipartite lattices},\ }\href
  {https://doi.org/10.1103/PhysRevX.12.021058} {\bibfield  {journal} {\bibinfo
  {journal} {Phys. Rev. X}\ }\textbf {\bibinfo {volume} {12}},\ \bibinfo
  {pages} {021058} (\bibinfo {year} {2022})}\BibitemShut {NoStop}%
\bibitem [{\citenamefont {Bhola}\ and\ \citenamefont
  {Damle}(2023)}]{Bhola_Damle_arXivOct2023}%
  \BibitemOpen
  \bibfield  {author} {\bibinfo {author} {\bibfnamefont {R.}~\bibnamefont
  {Bhola}}\ and\ \bibinfo {author} {\bibfnamefont {K.}~\bibnamefont {Damle}},\
  }\href@noop {} {\bibinfo {title} {{Gallai-Edmonds} percolation of
  topologically protected collective {M}ajorana excitations}} (\bibinfo {year}
  {2023}),\ \Eprint {https://arxiv.org/abs/2311.05634} {arXiv:2311.05634
  [cond-mat.dis-nn]} \BibitemShut {NoStop}%
\bibitem [{\citenamefont {Levin}\ and\ \citenamefont
  {Senthil}(2004)}]{Levin_Senthil_PRB_2004}%
  \BibitemOpen
  \bibfield  {author} {\bibinfo {author} {\bibfnamefont {M.}~\bibnamefont
  {Levin}}\ and\ \bibinfo {author} {\bibfnamefont {T.}~\bibnamefont
  {Senthil}},\ }\bibfield  {title} {\bibinfo {title} {Deconfined quantum
  criticality and {N\'eel} order via dimer disorder},\ }\href
  {https://doi.org/10.1103/PhysRevB.70.220403} {\bibfield  {journal} {\bibinfo
  {journal} {Phys. Rev. B}\ }\textbf {\bibinfo {volume} {70}},\ \bibinfo
  {pages} {220403} (\bibinfo {year} {2004})}\BibitemShut {NoStop}%
\bibitem [{\citenamefont {Senthil}\ \emph {et~al.}(2005)\citenamefont
  {Senthil}, \citenamefont {Balents}, \citenamefont {Sachdev}, \citenamefont
  {Vishwanath},\ and\ \citenamefont
  {P.~A.~Fisher}}]{Senthil_Balents_Sachdev_Vishwanath_Journal_Physical_Society_of_Japan_2005}%
  \BibitemOpen
  \bibfield  {author} {\bibinfo {author} {\bibfnamefont {T.}~\bibnamefont
  {Senthil}}, \bibinfo {author} {\bibfnamefont {L.}~\bibnamefont {Balents}},
  \bibinfo {author} {\bibfnamefont {S.}~\bibnamefont {Sachdev}}, \bibinfo
  {author} {\bibfnamefont {A.}~\bibnamefont {Vishwanath}},\ and\ \bibinfo
  {author} {\bibfnamefont {M.}~\bibnamefont {P.~A.~Fisher}},\ }\bibfield
  {title} {\bibinfo {title} {Deconfined criticality critically defined},\
  }\href {https://doi.org/10.1143/JPSJS.74S.1} {\bibfield  {journal} {\bibinfo
  {journal} {Journal of the Physical Society of Japan}\ }\textbf {\bibinfo
  {volume} {74}},\ \bibinfo {pages} {1} (\bibinfo {year} {2005})}\BibitemShut
  {NoStop}%
\bibitem [{\citenamefont {Affleck}(1985)}]{Affleck_PRL_1985}%
  \BibitemOpen
  \bibfield  {author} {\bibinfo {author} {\bibfnamefont {I.}~\bibnamefont
  {Affleck}},\ }\bibfield  {title} {\bibinfo {title} {Large-{$n$} limit of
  {$\mathrm{SU}(n)$} quantum "spin" chains},\ }\href
  {https://doi.org/10.1103/PhysRevLett.54.966} {\bibfield  {journal} {\bibinfo
  {journal} {Phys. Rev. Lett.}\ }\textbf {\bibinfo {volume} {54}},\ \bibinfo
  {pages} {966} (\bibinfo {year} {1985})}\BibitemShut {NoStop}%
\bibitem [{\citenamefont {Kaul}(2015)}]{Kaul_PRL_2015}%
  \BibitemOpen
  \bibfield  {author} {\bibinfo {author} {\bibfnamefont {R.~K.}\ \bibnamefont
  {Kaul}},\ }\bibfield  {title} {\bibinfo {title} {Spin nematics, valence-bond
  solids, and spin liquids in {$\mathrm{SO}(N)$} quantum spin models on the
  triangular lattice},\ }\href {https://doi.org/10.1103/PhysRevLett.115.157202}
  {\bibfield  {journal} {\bibinfo  {journal} {Phys. Rev. Lett.}\ }\textbf
  {\bibinfo {volume} {115}},\ \bibinfo {pages} {157202} (\bibinfo {year}
  {2015})}\BibitemShut {NoStop}%
\bibitem [{\citenamefont {Block}\ \emph {et~al.}(2020)\citenamefont {Block},
  \citenamefont {D'Emidio},\ and\ \citenamefont
  {Kaul}}]{Block_DEmidio_Kaul_PRB_2020}%
  \BibitemOpen
  \bibfield  {author} {\bibinfo {author} {\bibfnamefont {M.~S.}\ \bibnamefont
  {Block}}, \bibinfo {author} {\bibfnamefont {J.}~\bibnamefont {D'Emidio}},\
  and\ \bibinfo {author} {\bibfnamefont {R.~K.}\ \bibnamefont {Kaul}},\
  }\bibfield  {title} {\bibinfo {title} {Kagome model for a
  {${\mathbb{Z}}_{2}$} quantum spin liquid},\ }\href
  {https://doi.org/10.1103/PhysRevB.101.020402} {\bibfield  {journal} {\bibinfo
   {journal} {Phys. Rev. B}\ }\textbf {\bibinfo {volume} {101}},\ \bibinfo
  {pages} {020402} (\bibinfo {year} {2020})}\BibitemShut {NoStop}%
\bibitem [{\citenamefont {Harada}\ \emph {et~al.}(2003)\citenamefont {Harada},
  \citenamefont {Kawashima},\ and\ \citenamefont
  {Troyer}}]{Harada_Kawashima_Troyer_PRL_2003}%
  \BibitemOpen
  \bibfield  {author} {\bibinfo {author} {\bibfnamefont {K.}~\bibnamefont
  {Harada}}, \bibinfo {author} {\bibfnamefont {N.}~\bibnamefont {Kawashima}},\
  and\ \bibinfo {author} {\bibfnamefont {M.}~\bibnamefont {Troyer}},\
  }\bibfield  {title} {\bibinfo {title} {N\'eel and spin-peierls ground states
  of two-dimensional {$\mathrm{S}\mathrm{U}(N)$} quantum antiferromagnets},\
  }\href {https://doi.org/10.1103/PhysRevLett.90.117203} {\bibfield  {journal}
  {\bibinfo  {journal} {Phys. Rev. Lett.}\ }\textbf {\bibinfo {volume} {90}},\
  \bibinfo {pages} {117203} (\bibinfo {year} {2003})}\BibitemShut {NoStop}%
\bibitem [{\citenamefont {Kawashima}\ and\ \citenamefont
  {Tanabe}(2007)}]{Kawashima_Naoki_Tanabe_PRL_2007}%
  \BibitemOpen
  \bibfield  {author} {\bibinfo {author} {\bibfnamefont {N.}~\bibnamefont
  {Kawashima}}\ and\ \bibinfo {author} {\bibfnamefont {Y.}~\bibnamefont
  {Tanabe}},\ }\bibfield  {title} {\bibinfo {title} {Ground states of the
  {$\mathrm{SU}(N)$ Heisenberg} model},\ }\href
  {https://doi.org/10.1103/PhysRevLett.98.057202} {\bibfield  {journal}
  {\bibinfo  {journal} {Phys. Rev. Lett.}\ }\textbf {\bibinfo {volume} {98}},\
  \bibinfo {pages} {057202} (\bibinfo {year} {2007})}\BibitemShut {NoStop}%
\bibitem [{\citenamefont {Beach}\ \emph {et~al.}(2009)\citenamefont {Beach},
  \citenamefont {Alet}, \citenamefont {Mambrini},\ and\ \citenamefont
  {Capponi}}]{Beach_Alet_Mambrini_Capponi_PRB_2009}%
  \BibitemOpen
  \bibfield  {author} {\bibinfo {author} {\bibfnamefont {K.~S.~D.}\
  \bibnamefont {Beach}}, \bibinfo {author} {\bibfnamefont {F.}~\bibnamefont
  {Alet}}, \bibinfo {author} {\bibfnamefont {M.}~\bibnamefont {Mambrini}},\
  and\ \bibinfo {author} {\bibfnamefont {S.}~\bibnamefont {Capponi}},\
  }\bibfield  {title} {\bibinfo {title} {{$ \text{SU}(N)$} {H}eisenberg model
  on the square lattice: A continuous-{$N$} quantum {Monte Carlo} study},\
  }\href {https://doi.org/10.1103/PhysRevB.80.184401} {\bibfield  {journal}
  {\bibinfo  {journal} {Phys. Rev. B}\ }\textbf {\bibinfo {volume} {80}},\
  \bibinfo {pages} {184401} (\bibinfo {year} {2009})}\BibitemShut {NoStop}%
\bibitem [{\citenamefont {Lou}\ \emph {et~al.}(2009)\citenamefont {Lou},
  \citenamefont {Sandvik},\ and\ \citenamefont
  {Kawashima}}]{Lou_Sandvik_Kawashima_PRB_2009}%
  \BibitemOpen
  \bibfield  {author} {\bibinfo {author} {\bibfnamefont {J.}~\bibnamefont
  {Lou}}, \bibinfo {author} {\bibfnamefont {A.~W.}\ \bibnamefont {Sandvik}},\
  and\ \bibinfo {author} {\bibfnamefont {N.}~\bibnamefont {Kawashima}},\
  }\bibfield  {title} {\bibinfo {title} {Antiferromagnetic to
  valence-bond-solid transitions in two-dimensional {$ \text{SU}(N)$}
  {Heisenberg} models with multispin interactions},\ }\href
  {https://doi.org/10.1103/PhysRevB.80.180414} {\bibfield  {journal} {\bibinfo
  {journal} {Phys. Rev. B}\ }\textbf {\bibinfo {volume} {80}},\ \bibinfo
  {pages} {180414} (\bibinfo {year} {2009})}\BibitemShut {NoStop}%
\bibitem [{\citenamefont {Kaul}\ and\ \citenamefont
  {Sandvik}(2012)}]{Kaul_Sandvik_PRL_2012}%
  \BibitemOpen
  \bibfield  {author} {\bibinfo {author} {\bibfnamefont {R.~K.}\ \bibnamefont
  {Kaul}}\ and\ \bibinfo {author} {\bibfnamefont {A.~W.}\ \bibnamefont
  {Sandvik}},\ }\bibfield  {title} {\bibinfo {title} {Lattice model for the
  {$\mathrm{SU}(N)$} {N}\'eel to valence-bond solid quantum phase transition at
  large {$N$}},\ }\href {https://doi.org/10.1103/PhysRevLett.108.137201}
  {\bibfield  {journal} {\bibinfo  {journal} {Phys. Rev. Lett.}\ }\textbf
  {\bibinfo {volume} {108}},\ \bibinfo {pages} {137201} (\bibinfo {year}
  {2012})}\BibitemShut {NoStop}%
\bibitem [{\citenamefont {Block}\ \emph {et~al.}(2013)\citenamefont {Block},
  \citenamefont {Melko},\ and\ \citenamefont
  {Kaul}}]{Block_Melko_Kaul_PRL_2013}%
  \BibitemOpen
  \bibfield  {author} {\bibinfo {author} {\bibfnamefont {M.~S.}\ \bibnamefont
  {Block}}, \bibinfo {author} {\bibfnamefont {R.~G.}\ \bibnamefont {Melko}},\
  and\ \bibinfo {author} {\bibfnamefont {R.~K.}\ \bibnamefont {Kaul}},\
  }\bibfield  {title} {\bibinfo {title} {Fate of
  {$\mathbb{C}{\mathbb{P}}^{N\ensuremath{-}1}$} fixed points with $q$
  monopoles},\ }\href {https://doi.org/10.1103/PhysRevLett.111.137202}
  {\bibfield  {journal} {\bibinfo  {journal} {Phys. Rev. Lett.}\ }\textbf
  {\bibinfo {volume} {111}},\ \bibinfo {pages} {137202} (\bibinfo {year}
  {2013})}\BibitemShut {NoStop}%
\bibitem [{\citenamefont {Kundu}\ \emph {et~al.}(2023)\citenamefont {Kundu},
  \citenamefont {Desai},\ and\ \citenamefont
  {Damle}}]{Kundu_Desai_Damle_arXivSep2023}%
  \BibitemOpen
  \bibfield  {author} {\bibinfo {author} {\bibfnamefont {S.}~\bibnamefont
  {Kundu}}, \bibinfo {author} {\bibfnamefont {N.}~\bibnamefont {Desai}},\ and\
  \bibinfo {author} {\bibfnamefont {K.}~\bibnamefont {Damle}},\ }\href@noop {}
  {\bibinfo {title} {Competition between {Neel}, {Haldane nematic, plaquette
  valence bond solid, and $(\pi,\pi)$ valence bond solid phases in SU(N)
  analogs of $S=1$ square-lattice antiferromagnets}}} (\bibinfo {year}
  {2023}),\ \Eprint {https://arxiv.org/abs/2309.12262} {arXiv:2309.12262
  [cond-mat.str-el]} \BibitemShut {NoStop}%
\bibitem [{\citenamefont {Sandvik}(2007)}]{Sandvik_PRL_2007}%
  \BibitemOpen
  \bibfield  {author} {\bibinfo {author} {\bibfnamefont {A.~W.}\ \bibnamefont
  {Sandvik}},\ }\bibfield  {title} {\bibinfo {title} {Evidence for deconfined
  quantum criticality in a two-dimensional {Heisenberg} model with four-spin
  interactions},\ }\href {https://doi.org/10.1103/PhysRevLett.98.227202}
  {\bibfield  {journal} {\bibinfo  {journal} {Phys. Rev. Lett.}\ }\textbf
  {\bibinfo {volume} {98}},\ \bibinfo {pages} {227202} (\bibinfo {year}
  {2007})}\BibitemShut {NoStop}%
\bibitem [{\citenamefont {Melko}\ and\ \citenamefont
  {Kaul}(2008)}]{Melko_Kaul_PRL_2008}%
  \BibitemOpen
  \bibfield  {author} {\bibinfo {author} {\bibfnamefont {R.~G.}\ \bibnamefont
  {Melko}}\ and\ \bibinfo {author} {\bibfnamefont {R.~K.}\ \bibnamefont
  {Kaul}},\ }\bibfield  {title} {\bibinfo {title} {Scaling in the fan of an
  unconventional quantum critical point},\ }\href
  {https://doi.org/10.1103/PhysRevLett.100.017203} {\bibfield  {journal}
  {\bibinfo  {journal} {Phys. Rev. Lett.}\ }\textbf {\bibinfo {volume} {100}},\
  \bibinfo {pages} {017203} (\bibinfo {year} {2008})}\BibitemShut {NoStop}%
\bibitem [{\citenamefont {Sandvik}(2010{\natexlab{a}})}]{Sandvik_PRL_2010}%
  \BibitemOpen
  \bibfield  {author} {\bibinfo {author} {\bibfnamefont {A.~W.}\ \bibnamefont
  {Sandvik}},\ }\bibfield  {title} {\bibinfo {title} {Continuous quantum phase
  transition between an antiferromagnet and a valence-bond solid in two
  dimensions: Evidence for logarithmic corrections to scaling},\ }\href
  {https://doi.org/10.1103/PhysRevLett.104.177201} {\bibfield  {journal}
  {\bibinfo  {journal} {Phys. Rev. Lett.}\ }\textbf {\bibinfo {volume} {104}},\
  \bibinfo {pages} {177201} (\bibinfo {year} {2010}{\natexlab{a}})}\BibitemShut
  {NoStop}%
\bibitem [{\citenamefont {Sen}\ and\ \citenamefont
  {Sandvik}(2010)}]{Sen_Sandvik_PRB_2010}%
  \BibitemOpen
  \bibfield  {author} {\bibinfo {author} {\bibfnamefont {A.}~\bibnamefont
  {Sen}}\ and\ \bibinfo {author} {\bibfnamefont {A.~W.}\ \bibnamefont
  {Sandvik}},\ }\bibfield  {title} {\bibinfo {title} {Example of a first-order
  {N\'eel} to valence-bond-solid transition in two dimensions},\ }\href
  {https://doi.org/10.1103/PhysRevB.82.174428} {\bibfield  {journal} {\bibinfo
  {journal} {Phys. Rev. B}\ }\textbf {\bibinfo {volume} {82}},\ \bibinfo
  {pages} {174428} (\bibinfo {year} {2010})}\BibitemShut {NoStop}%
\bibitem [{\citenamefont {Banerjee}\ \emph
  {et~al.}(2011{\natexlab{b}})\citenamefont {Banerjee}, \citenamefont {Damle},\
  and\ \citenamefont {Paramekanti}}]{Banerjee_Damle_Paramekanti_PRB_2011}%
  \BibitemOpen
  \bibfield  {author} {\bibinfo {author} {\bibfnamefont {A.}~\bibnamefont
  {Banerjee}}, \bibinfo {author} {\bibfnamefont {K.}~\bibnamefont {Damle}},\
  and\ \bibinfo {author} {\bibfnamefont {A.}~\bibnamefont {Paramekanti}},\
  }\bibfield  {title} {\bibinfo {title} {N\'eel to staggered dimer order
  transition in a generalized honeycomb lattice {Heisenberg model}},\ }\href
  {https://doi.org/10.1103/PhysRevB.83.134419} {\bibfield  {journal} {\bibinfo
  {journal} {Phys. Rev. B}\ }\textbf {\bibinfo {volume} {83}},\ \bibinfo
  {pages} {134419} (\bibinfo {year} {2011}{\natexlab{b}})}\BibitemShut
  {NoStop}%
\bibitem [{\citenamefont {Pujari}\ \emph {et~al.}(2013)\citenamefont {Pujari},
  \citenamefont {Damle},\ and\ \citenamefont
  {Alet}}]{Pujari_Damle_Alet_PRL_2013}%
  \BibitemOpen
  \bibfield  {author} {\bibinfo {author} {\bibfnamefont {S.}~\bibnamefont
  {Pujari}}, \bibinfo {author} {\bibfnamefont {K.}~\bibnamefont {Damle}},\ and\
  \bibinfo {author} {\bibfnamefont {F.}~\bibnamefont {Alet}},\ }\bibfield
  {title} {\bibinfo {title} {N\'eel-state to valence-bond-solid transition on
  the honeycomb lattice: Evidence for deconfined criticality},\ }\href
  {https://doi.org/10.1103/PhysRevLett.111.087203} {\bibfield  {journal}
  {\bibinfo  {journal} {Phys. Rev. Lett.}\ }\textbf {\bibinfo {volume} {111}},\
  \bibinfo {pages} {087203} (\bibinfo {year} {2013})}\BibitemShut {NoStop}%
\bibitem [{\citenamefont {Pujari}\ \emph {et~al.}(2015)\citenamefont {Pujari},
  \citenamefont {Alet},\ and\ \citenamefont
  {Damle}}]{Pujari_Alet_Damle_PRB_2015}%
  \BibitemOpen
  \bibfield  {author} {\bibinfo {author} {\bibfnamefont {S.}~\bibnamefont
  {Pujari}}, \bibinfo {author} {\bibfnamefont {F.}~\bibnamefont {Alet}},\ and\
  \bibinfo {author} {\bibfnamefont {K.}~\bibnamefont {Damle}},\ }\bibfield
  {title} {\bibinfo {title} {Transitions to valence-bond solid order in a
  honeycomb lattice antiferromagnet},\ }\href
  {https://doi.org/10.1103/PhysRevB.91.104411} {\bibfield  {journal} {\bibinfo
  {journal} {Phys. Rev. B}\ }\textbf {\bibinfo {volume} {91}},\ \bibinfo
  {pages} {104411} (\bibinfo {year} {2015})}\BibitemShut {NoStop}%
\bibitem [{\citenamefont {Iaizzi}\ \emph {et~al.}(2017)\citenamefont {Iaizzi},
  \citenamefont {Damle},\ and\ \citenamefont
  {Sandvik}}]{Iaizzi_Damle_Sandvik_PRB_2017}%
  \BibitemOpen
  \bibfield  {author} {\bibinfo {author} {\bibfnamefont {A.}~\bibnamefont
  {Iaizzi}}, \bibinfo {author} {\bibfnamefont {K.}~\bibnamefont {Damle}},\ and\
  \bibinfo {author} {\bibfnamefont {A.~W.}\ \bibnamefont {Sandvik}},\
  }\bibfield  {title} {\bibinfo {title} {Field-driven quantum phase transitions
  in {$S=\frac{1}{2}$} spin chains},\ }\href
  {https://doi.org/10.1103/PhysRevB.95.174436} {\bibfield  {journal} {\bibinfo
  {journal} {Phys. Rev. B}\ }\textbf {\bibinfo {volume} {95}},\ \bibinfo
  {pages} {174436} (\bibinfo {year} {2017})}\BibitemShut {NoStop}%
\bibitem [{\citenamefont {Iaizzi}\ \emph {et~al.}(2018)\citenamefont {Iaizzi},
  \citenamefont {Damle},\ and\ \citenamefont
  {Sandvik}}]{Iaizzi_Damle_Sandvik_PRB_2018}%
  \BibitemOpen
  \bibfield  {author} {\bibinfo {author} {\bibfnamefont {A.}~\bibnamefont
  {Iaizzi}}, \bibinfo {author} {\bibfnamefont {K.}~\bibnamefont {Damle}},\ and\
  \bibinfo {author} {\bibfnamefont {A.~W.}\ \bibnamefont {Sandvik}},\
  }\bibfield  {title} {\bibinfo {title} {Metamagnetism and zero-scale-factor
  universality in the two-dimensional {$J\text{\ensuremath{-}}Q$} model},\
  }\href {https://doi.org/10.1103/PhysRevB.98.064405} {\bibfield  {journal}
  {\bibinfo  {journal} {Phys. Rev. B}\ }\textbf {\bibinfo {volume} {98}},\
  \bibinfo {pages} {064405} (\bibinfo {year} {2018})}\BibitemShut {NoStop}%
\bibitem [{\citenamefont {Kececioglu}\ and\ \citenamefont
  {Pecqueur}(1998)}]{Kececioglu_Pecqueur_MaxMatch_1998}%
  \BibitemOpen
  \bibfield  {author} {\bibinfo {author} {\bibfnamefont {J.~D.}\ \bibnamefont
  {Kececioglu}}\ and\ \bibinfo {author} {\bibfnamefont {J.}~\bibnamefont
  {Pecqueur}},\ }\bibfield  {title} {\bibinfo {title} {Computing
  maximum-cardinality matchings in sparse general graphs},\ }in\ \href
  {https://api.semanticscholar.org/CorpusID:9752229} {\emph {\bibinfo
  {booktitle} {Workshop on Algorithm Engineering}}}\ (\bibinfo {year} {1998})\
  pp.\ \bibinfo {pages} {121--132}\BibitemShut {NoStop}%
\bibitem [{\citenamefont {Sandvik}(1992)}]{Sandvik_JPhysA_1992}%
  \BibitemOpen
  \bibfield  {author} {\bibinfo {author} {\bibfnamefont {A.~W.}\ \bibnamefont
  {Sandvik}},\ }\bibfield  {title} {\bibinfo {title} {A generalization of
  {H}andscomb's quantum {Monte Carlo scheme-application to the 1D Hubbard
  model}},\ }\href {https://doi.org/10.1088/0305-4470/25/13/017} {\bibfield
  {journal} {\bibinfo  {journal} {Journal of Physics A: Mathematical and
  General}\ }\textbf {\bibinfo {volume} {25}},\ \bibinfo {pages} {3667}
  (\bibinfo {year} {1992})}\BibitemShut {NoStop}%
\bibitem [{\citenamefont {Sandvik}(1999)}]{Sandvik_PRB_1999}%
  \BibitemOpen
  \bibfield  {author} {\bibinfo {author} {\bibfnamefont {A.~W.}\ \bibnamefont
  {Sandvik}},\ }\bibfield  {title} {\bibinfo {title} {Stochastic series
  expansion method with operator-loop update},\ }\href
  {https://doi.org/10.1103/PhysRevB.59.R14157} {\bibfield  {journal} {\bibinfo
  {journal} {Phys. Rev. B}\ }\textbf {\bibinfo {volume} {59}},\ \bibinfo
  {pages} {R14157} (\bibinfo {year} {1999})}\BibitemShut {NoStop}%
\bibitem [{\citenamefont {Sylju\aa{}sen}\ and\ \citenamefont
  {Sandvik}(2002)}]{Syljuasen_Sandvik_PRE_2002}%
  \BibitemOpen
  \bibfield  {author} {\bibinfo {author} {\bibfnamefont {O.~F.}\ \bibnamefont
  {Sylju\aa{}sen}}\ and\ \bibinfo {author} {\bibfnamefont {A.~W.}\ \bibnamefont
  {Sandvik}},\ }\bibfield  {title} {\bibinfo {title} {Quantum {M}onte {C}arlo
  with directed loops},\ }\href {https://doi.org/10.1103/PhysRevE.66.046701}
  {\bibfield  {journal} {\bibinfo  {journal} {Phys. Rev. E}\ }\textbf {\bibinfo
  {volume} {66}},\ \bibinfo {pages} {046701} (\bibinfo {year}
  {2002})}\BibitemShut {NoStop}%
\bibitem [{\citenamefont
  {Sandvik}(2010{\natexlab{b}})}]{Sandvik_AIP_Conference_2010}%
  \BibitemOpen
  \bibfield  {author} {\bibinfo {author} {\bibfnamefont {A.~W.}\ \bibnamefont
  {Sandvik}},\ }\bibfield  {title} {\bibinfo {title} {Computational studies of
  quantum spin systems},\ }\href {https://doi.org/10.1063/1.3518900} {\bibfield
   {journal} {\bibinfo  {journal} {AIP Conference Proceedings}\ }\textbf
  {\bibinfo {volume} {1297}},\ \bibinfo {pages} {135} (\bibinfo {year}
  {2010}{\natexlab{b}})}\BibitemShut {NoStop}%
\bibitem [{\citenamefont {Sandvik}\ and\ \citenamefont
  {Evertz}(2010)}]{Sandvik_Evertz_PRB_2010}%
  \BibitemOpen
  \bibfield  {author} {\bibinfo {author} {\bibfnamefont {A.~W.}\ \bibnamefont
  {Sandvik}}\ and\ \bibinfo {author} {\bibfnamefont {H.~G.}\ \bibnamefont
  {Evertz}},\ }\bibfield  {title} {\bibinfo {title} {Loop updates for
  variational and projector quantum {Monte Carlo} simulations in the
  valence-bond basis},\ }\href {https://doi.org/10.1103/PhysRevB.82.024407}
  {\bibfield  {journal} {\bibinfo  {journal} {Phys. Rev. B}\ }\textbf {\bibinfo
  {volume} {82}},\ \bibinfo {pages} {024407} (\bibinfo {year}
  {2010})}\BibitemShut {NoStop}%
\bibitem [{\citenamefont {Banerjee}\ and\ \citenamefont
  {Damle}(2010)}]{Banerjee_Damle_JStatMech_2010}%
  \BibitemOpen
  \bibfield  {author} {\bibinfo {author} {\bibfnamefont {A.}~\bibnamefont
  {Banerjee}}\ and\ \bibinfo {author} {\bibfnamefont {K.}~\bibnamefont
  {Damle}},\ }\bibfield  {title} {\bibinfo {title} {Generalization of the
  singlet sector valence-bond loop algorithm to antiferromagnetic ground states
  with total spin {$\rm S_{tot}$} = 1/2},\ }\href
  {https://doi.org/10.1088/1742-5468/2010/08/P08017} {\bibfield  {journal}
  {\bibinfo  {journal} {Journal of Statistical Mechanics: Theory and
  Experiment}\ }\textbf {\bibinfo {volume} {2010}},\ \bibinfo {pages} {P08017}
  (\bibinfo {year} {2010})}\BibitemShut {NoStop}%
\bibitem [{\citenamefont {Desai}\ and\ \citenamefont
  {Pujari}(2021)}]{Desai_Pujari_PRB_2021}%
  \BibitemOpen
  \bibfield  {author} {\bibinfo {author} {\bibfnamefont {N.}~\bibnamefont
  {Desai}}\ and\ \bibinfo {author} {\bibfnamefont {S.}~\bibnamefont {Pujari}},\
  }\bibfield  {title} {\bibinfo {title} {Resummation-based quantum {Monte
  Carlo} for quantum paramagnetic phases},\ }\href
  {https://doi.org/10.1103/PhysRevB.104.L060406} {\bibfield  {journal}
  {\bibinfo  {journal} {Phys. Rev. B}\ }\textbf {\bibinfo {volume} {104}},\
  \bibinfo {pages} {L060406} (\bibinfo {year} {2021})}\BibitemShut {NoStop}%
\end{thebibliography}%

\end{document}